%% file: main.tex
\documentclass[journal=jpcafh,manuscript=article]{achemso}

\usepackage[version=3]{mhchem} 
\usepackage[utf8]{inputenc}
\usepackage[T1]{fontenc}
\usepackage{lmodern}
\usepackage{amsmath}
\usepackage{amsfonts}
\usepackage{amssymb}
\usepackage{graphicx}
\usepackage{textgreek}

\usepackage{subfig}
\usepackage{pdfpages}
\usepackage{booktabs}
\newcommand{\tabitem}{~~\llap{\textbullet}~~}
\usepackage{geometry}
\usepackage{mathtools}
\usepackage{colortbl}
\usepackage{pdflscape}
\usepackage{adjustbox}
\usepackage{lipsum}
\usepackage{colortbl}
\usepackage{xcolor}
\definecolor{myred}{RGB}{255, 153, 153} 
\definecolor{myblue}{RGB}{204, 204, 255}
\definecolor{mygreen}{RGB}{153, 255, 153}
\definecolor{mypurple}{RGB}{204, 153, 255}

\include{macros}

\author{Sehr Naseem-Khan}
\affiliation[SorbonneU]{Laboratoire de Chimie Théorique, Sorbonne Université, UMR 7616 CNRS, 75005, Paris, France}
\author{Nohad Gresh}
\affiliation[SorbonneU]{Laboratoire de Chimie Théorique, Sorbonne Université, UMR 7616 CNRS, 75005, Paris, France}
\author{Alston J. Misquitta}
\affiliation[Queen Mary]
{School of Physics and Astronomy and the Thomas Young Centre for Theory and Simulation of
Materials at Queen Mary University of London, London E1 4NS, U.K.}
\email{a.j.misquitta@qmul.ac.uk}
\author{Jean-Philip Piquemal}
\affiliation[SorbonneU]
{Laboratoire de Chimie Théorique, Sorbonne Université, UMR 7616 CNRS, 75005, Paris, France}
\alsoaffiliation[IUF]{Institut Universitaire de France, 75005, Paris, France}
\alsoaffiliation[UT]{The University of Texas at Austin, Department of Biomedical Engineering, TX, USA}
\email{jean-philip.piquemal@sorbonne-universite.fr}

\title[SAPT-Article]{Assessment of SAPT and Supermolecular EDAs Approaches for the Development of Separable and Polarizable force fields}

\begin{document}

\begin{tocentry}
\begin{center}
    \includegraphics[scale=0.18]{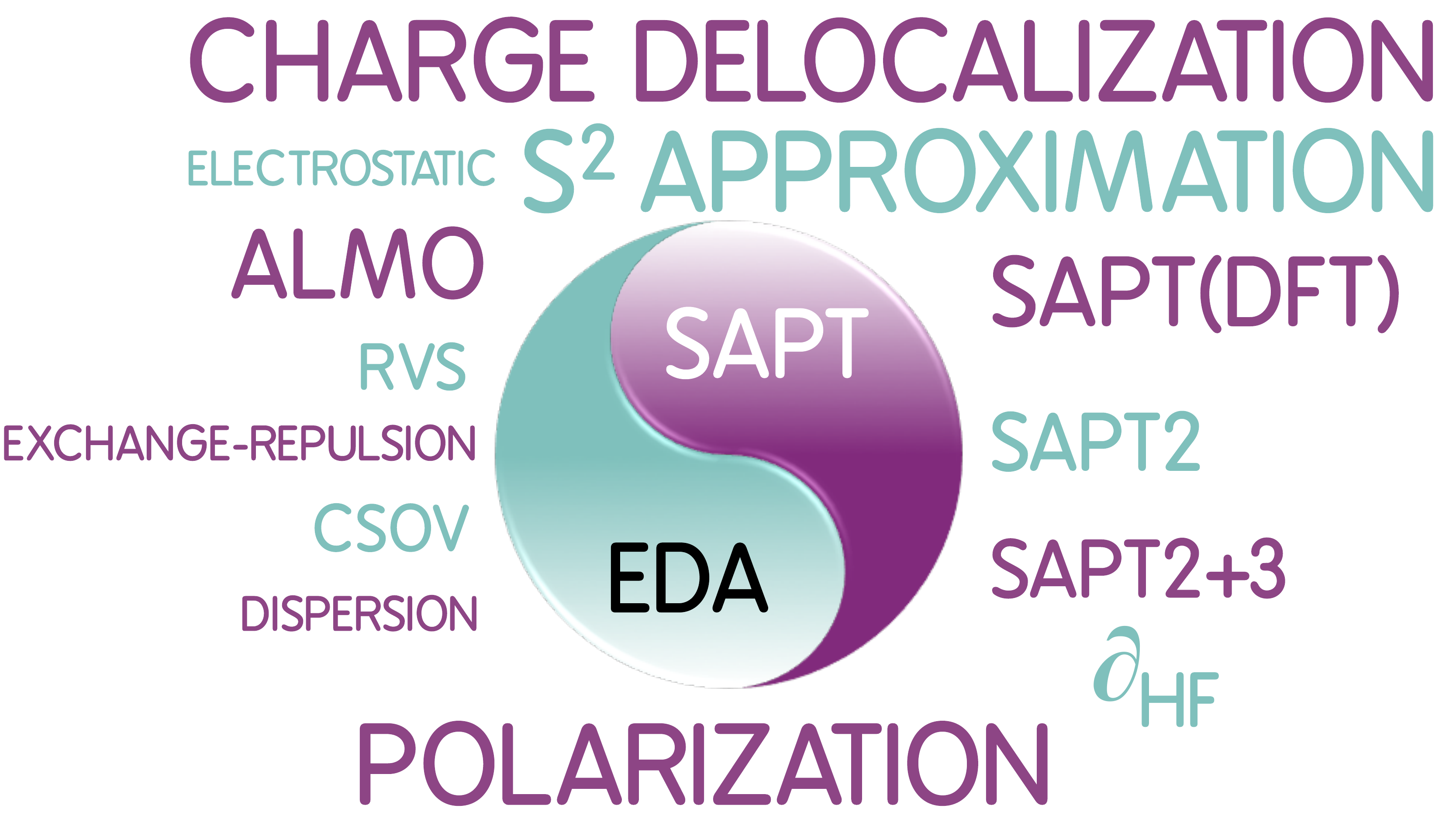}  
\end{center}
  




\end{tocentry}


\begin{abstract}

Which is the best reference quantum chemical approach to decipher the energy components of the total interaction energy : Symmetry-Adapted Perturbation Theory (\sapt) or Supermolecular Energy Decomposition Analysis (EDA) methods? With the rise of physically motivated polarizable force fields (\polFF) grounded on these procedures, the need to answer such a question becomes critical. We report a systematic and detailed assessment of three variants of \sapt (namely \SAPT{2}, \SAPT{2+3} and \saptdft) and three supermolecular EDAs approaches (\almo, \csov and \rvs). A set of challenging, strongly bound water complexes: \waterdimer, \Znwater and \Fwater are used as ‘stress-tests’ for these electronic structure methods. We have developed a procedure to separate the induction energy into the polarization and charge-delocalization using an infinite-order strategy based on \saptdft. This paper aims to provide an overview of the capabilities and limitations, but also similarities, of \sapt and supermolecular EDAs approaches for \polFF developments. Our results show that \saptdftnossq and \almofunc{\wbxd} are the most accurate and reliable techniques.


\end{abstract}



\section{Introduction}
The development of accurate intermolecular potentials is critical for the production of robust and predictive molecular dynamics simulations. These potentials commonly rely on gas phase quantum-chemical data, but sometime can also be combined with condensed phase experimental results as a reference for their parametrization. While emerging techniques such as machine learning tend to fit a large amount of data to provide a black-box prediction of the total intermolecular interaction energy (\EINT), the development of polarizable force fields (\polFF) follows a physics-based strategy. In these approaches, \EINT is reconstructed upon computing a sum of physically motivated contributions such as: electrostatics, exchange-repulsion, many-body induction including polarization and charge-transfer (charge-delocalization), and dispersion. Of course, such a strategy requires to go beyond classical force fields definitions and to define more complex functional forms for each component of the interaction potential. However, on a Chemistry point of view, the separable aspect of these potentials (i.e. their ability to reconstruct \EINT contribution by contribution) brings additional and powerful interpretative capabilities to users as information on the physical origins of the molecular interactions is automatically provided.
In our groups, we favour this latter approach.\cite{GEM0,NohadGresh2007,reviewcompchem, annurev-biophys-070317-033349,melcr2019accurate,MisquittaS16,liu2019amoeba+,GilmoreDM_water_2020} There are now numerous theoretical ways of computing these terms, but there is, as yet, no direct means of experimental validation of all energy components \cite{andres_nine_questions_2019}. Indeed, Experiment essentially captures long-range quantities such as multipole moments and frequency-dependent polarizabilities (See Stone's book, Ch. 13) \cite{Stone:book:13}. Alternatively, one could try to resort to indirect means of assessment to correlate the theoretical predictions to experimental measurements such as bond lengths or spectra \cite{FrenkingEDA,Mobenzene} but it becomes rapidly tedious when a general force field, i.e. valid for any type of element,  has to be derived. For these reasons, the gas phase parametrization of intermolecular potentials remains essentially grounded on theoretical results. It is important to note that such \textit{ab initio} force fields should be transferable enough to be able to conserve their initial parametrization to achieve condensed phase simulations. Therefore, the development of next generation accurate \polFF rises the key question of the critical choice of a reference technique able to decipher the physical components of \EINT. Such an approach should: i) possess an overall accuracy leading to \EINT values comparable to high level \abinitio reference data such as \ccsdt; ii) should have a robust numerical stability, satisfactory physical trends and a correct asymptotic behaviour for all its different energetic components. The final challenge here is to also obtain contributions summing up to an accurate and predictive value of \EINT.

The theoretical methods able to unravel \EINT energetic components can be separated into two classes of techniques: perturbational and supermolecular. The perturbational approaches compute \EINT as the sum of physical terms using some form of intermolecular perturbation theory, while the supermolecular strategies decompose a supermolecular interaction energy into physical components. 
Among the perturbational methods, the Symmetry-Adapted Perturbation Theory (\sapt) approach \cite{Jeziorski1994} is the most well-known and widely applied perturbational strategy to compute intermolecular interaction energies between molecular dimers and trimers.
Over the years, \sapt has evolved into families of techniques differing one from another based on the choice of the treatment of the intermolecular electronic correlation  \cite{BukowskiSJJSKWR99,misquitta2000spectra,Parker2014,korona2010coupled}. These different \sapt methods can now be directly accessed in different codes as {\sc SAPT2020} suite of codes\cite{garcia2020}, \PsiF package \cite{Psi41.1,Turney2012}, \Molpro \cite{MOLPRO-WIREs} and \CamCASP \cite{misquitta_camcasp}.
The {\sc SAPT2020} code and its earlier versions have been used in pioneering developments in intermolecular potentials for a variety of complexes including high-accuracy potentials for small complexes, or models for water embedding explicit two-body and three-body effects \cite{garcia2020}. In parallel, with the development of \saptdft \cite{Hebelmann2002,Hebelmann2002bis,Hebelmann2003, Misquitta2002,Misquitta2003,misquitta_saptdft_2005}, interaction potentials for small to medium-sized organic molecules have been developed (see for example the reviews by Szalewicz \cite{szalewicz_intermolecular_2005-1,szalewicz_symmetry-adapted_2012} and references therein). Likewise, Misquitta \& Stone have pioneered a series of accurate, many-body intermolecular interaction models for small organic molecules with \CamCASP \cite{misquitta_first_2008,Misquitta2016,GilmoreDM_water_2020}.  
However, both the number and diversity of applications of \sapt increased remarkably following its implementation in the \PsiF package. This was primarily because of the ease of use of \PsiF, but also because of the improvements to the computational efficiency of the implementation using techniques like density-fitting. Thus, electrostatic models have been developed based on \sapt within the framework of the AMOEBA \cite{wang2015general} and SIBFA \cite{Narth2016} polFFs in order to include charge penetration effects.\cite{emtp} 
\sapt has also been used to develop many-body analytical potential functions \cite{cisneros2016modeling,melcr2019accurate}, and advanced polarizable water models such as AMOEBA+\cite{liu2019amoeba+,liu2020amoeba+} and the Gaussian Electrostatic Model (GEM)\cite{GEM0,Duke2014,GEMstatus}.

On the other hand, supermolecular EDAs for the analysis of the intermolecular interaction energy pre-date the \sapt methods, and continue to be developed. Some of these encompass the Constrained Space Orbital Variation (\csov) \cite{Bagus1984}, Reduced Variational Space (\rvs) \cite{Stevens1987}, Block Localized Wave Function (BLW) \cite{mo2000,mo2007,mo2011} and Absolutely Localized Molecular Orbitals (\almo) \cite{Khaliullin2007} methods. They decompose \EINT into Coulomb (electrostatic) and exchange-repulsion energies at first-order, and polarization and charge-transfer (charge-delocalization) energies at second and higher-orders. \rvs and \csov have been used to demonstrate the importance of the charge-delocalization contribution in water and also in \Zn complexes.\cite{garmer1994,gresh1995mono,Piquemal2005} Recently, \almo was used to refine the \amoeba model for ion \dots water complexes \cite{mao2016assessing} and to develop the MB-UCB water model. \cite{Das2019} Also, supermolecular EDAs have enabled continuous refinements and validations of the SIBFA polarizable force-field. \cite{Gresh1986,gresh1995mono,Piquemal2003,Gresh2005,NohadGresh2007,Piquemal2007}

In this paper, we seek guidelines for the most robust methods enabling to study strongly polar and ionic complexes which are of critical importance for \polFF development. We will assess and compare perturbative and supermolecular approaches on the following complexes (\figrff{fig:structures}) : \waterdimer (neutral), \Znwater (cationic), \Fwater (anionic) and with supporting data for \Clwater and \OHwater provided in the \SI (Figure S1). The choice of \Zn is motivated by its strong ``polarizing'' divalent nature, its outstanding role in biochemistry being the second most abundant transition metal in organisms after iron, and as a structural element for Zn-fingers \cite{maynard2001reactivity} or as a co-factor \cite{lipscomb1996recent}. Due to its key importance in living organisms, \water is the first molecule considered for \polFF validation. And the ion \dots \water interactions are representative of the strongly bound, polar systems found in biological environments. An additional motivation to treat these systems is that they can be considered as `stress-tests' for electronic structure methods owing to the strong polarization and charge-transfer (charge-delocalization) effects at play these complexes.
This article is organized as follows: we first describe the \sapt, supermolecular EDAs methods and the polarization model developed in this work. We then assess the accuracy of \EINT from \sapt models and supermolecular EDAs with respect to \ccsdt, which is the `gold standard' method in quantum chemistry \cite{Rezac2013}. This is the first criterion to be validated for the choice of reference for \polFF development. 
We next investigate the effect of the single-exchange, or \Ssq, approximation \cite{Jeziorski1994} used to compute higher-order exchange terms in all \sapt methods. Sch\"affer \& Jansen \cite{SchafferJ12,SchafferJ_single-determinant-based_2013} have derived second-order exchange-induction and exchange-dispersion energy terms (at zeroth order in intramolecular correlation) without this approximation. They have demonstrated the significance of these new terms, particularly the second-order exchange-induction, for strongly bound systems.
We finally examine in detail the partitioning of the induction energy into the polarization and charge-delocalization. We demonstrate the numerical advantages of the approach based on regularized \saptdft \cite{Misquitta13}, and show how the \Ssq approximation has a significant impact on these energies. Also, using polarization models, we estimate the infinite-order polarization and charge-delocalization energies from \saptdft, and compare these to the supermolecular EDAs. We finally put forth a choice of \abinitio reference for \polFF development.

\input{Structures-Representation} 
\section{Methods}
\label{sec:methods}
The different supermolecular EDAs and \sapt methods used in this work are first described. Mathematical details can be found in Refs. \cite{Bagus1984,Stevens1987,Khaliullin2007,Jeziorski1994,Hebelmann2002,Hebelmann2002bis,Hebelmann2003,Misquitta2002,Misquitta2003} 
And a definition of charge-transfer (charge-delocalization) and polarization energies within each method is provided as well as a definition of the infinite-order energies from \saptdft.

\textit{Charge-delocalization (CD) or charge-transfer (CT)?}
Note that one of the contributions of the intermolecular interaction energy is associated with the sharing or tunneling of the electrons of the interacting monomers onto the electron-deficient sites of the partners, resulting in a lowering of the energy of the complex. This term is often termed as ``charge-transfer'' but following Misquitta, \cite{Misquitta13,GilmoreDM_water_2020} we will instead use a more appropriate term i.e. ``charge-delocalization'' (CD). While this may seem only a nomenclature issue, as discussed by Misquitta \cite{Misquitta13}, and also by Mao \etal \cite{Mao2018}, the term charge-transfer does not satisfactorily describe the process of electron sharing in a symmetric hydrogen-bonding dimer where there is no net charge transferred from one molecule to the other.

\subsection{Supermolecular EDAs}
\label{sec:EDAs}
Most EDAs resolve the supermolecular interaction energy into five contributions: electrostatic, exchange-repulsion, polarization, charge-delocalization and dispersion, but the decomposition schemes are not unique. 
A general expression of \EINT between two monomers A and B can be written as:
\begin{equation} \label{eq:sup-EDA}
    \EINT = \Ecomp{elst} + \Ecomp{exch-rep} + \Ecomp{pol} + \Ecomp{cd} + \Ecomp{disp}.
\end{equation}
The dispersion contribution is not to be included in the absence of electron correlation.
\rvs and \csov decompose \EINT using similar approaches. Indeed, the variational or the constrained spaces are divided into several sets of orbitals combining occupied and virtual orbitals of monomers A and B. Different constructions of these sets of orbitals allow to compute electrostatic, exchange-repulsion, polarization and charge-delocalization energies. In that case, the charge-delocalization energy is computed as the sum of electron delocalization energies from monomer A towards monomer B, and reciprocally, such as:
\begin{equation}
    \mathrm{E_{cd} = E_{cd(A\rightarrow B)} + E_{cd(B\rightarrow A)}}.
    \label{eq:CD}
\end{equation}
\rvs is limited to the Hartree--Fock (HF) level of theory. Thus, computing HF orbitals does not enable to take into account the correlation of opposite spin electrons. \csov was extended to the \dft level of theory, \cite{Piquemal2005,MARQUEZ1999463} but can also be used with multi-configurational SCF wavefunctions to access open-shell systems. \cite{Bauschlicher1986,MARJOLIN201325}

On the other hand, the \almo method decomposes \EINT at both HF and DFT levels of theory as:
\begin{equation} \label{eq:ALMO}
    \EINT^{\mathrm{ALMO}} = \Ecomp{frz} + \Ecomp{pol} + \Ecomp{cd},
\end{equation} 
where the \Ecomp{frz} term corresponds to the sum of electrostatic and exchange-repulsion energies computed with frozen (frz) orbitals.
The \almo method can be distinguished from \rvs and \csov through the use of absolutely localized molecular orbitals (ALMOs) that are expanded in terms of atomic orbitals (AOs) of a given molecule.\cite{Khaliullin2006} In other words, the ALMOs are centered on the atoms of the monomer as opposed to the MOs in \rvs or \csov which are delocalized over all the monomers.
Thus, the use of ALMOs prevents (or suppresses) intermolecular charge delocalization from one monomer to another monomer, allowing then a separation between polarization and charge delocalization terms. 
The CD-free state interaction energy, $E[\Psi_{\rm\almo}]$, is first computed with relaxed ALMOs, and the full interaction energy, $E[\Psi_{\rm full}]$, is subsequently computed between the fully optimized delocalized MOs, enabling \ecd to be derived as:
\begin{equation} \label{eq:ALMO-CD}
    \Ecomp{cd} = E[\Psi_{\rm full}] - E[\Psi_{\rm\almo}]. 
\end{equation}

The \rvs, \csov and \almo methods all remove the Basis Set Superposition Error (BSSE) within the charge delocalization contribution by using the counterpoise correction \cite{BSSE}. 

\subsection{\sapt methods}
\label{sec:SAPT}
Symmetry-Adapted Perturbation Theory (\sapt) is a class of intermolecular perturbation theories which are commonly based on symmetrized Rayleigh--Schr\"odinger (RS) perturbation theory.
It has become common to label \sapt and related methods as EDAs but this may not be appropriate since in \sapt, the interaction energy is built up, term-by-term, to give a total: there is no total energy that is partitioned as is done in supermolecular EDAs. Thus, \sapt based on Hartree--Fock orbitals, or SAPT(HF), is not an EDA for the Hartree--Fock-level interaction energy, even though it can be used to partition \EintHF if the dispersion term in SAPT(HF) is neglected. But it is a framework to construct the correlated interaction energy using the Hartree--Fock orbitals as a starting point for the perturbation expansion. Likewise, \saptdft starts from the Kohn-Sham orbitals and builds up a correlated interaction energy with a possibly further improved accuracy with respect to density-functional theory.

The interaction energy in \sapt can be expressed as a series expansion\cite{Jeziorski1994} as:
\begin{equation}
    \EintSAPT{} = \sum_{i=1}^\infty \sum_{j=0}^\infty \left( \EpolSAPT{ij} + \EexchSAPT{ij} \right),
        \label{eq:SAPT}
\end{equation}
where $i$ and $j$ indicate the order of intermolecular and intramolecular perturbation respectively. The so-called polarization term \EpolSAPT{ij} --- not to be confused with the polarization energy in a classical polarization model --- contains contributions from the electrostatic, induction and dispersion energies.
Each of these terms is associated a corresponding exchange term \EexchSAPT{ij} that arises from the (anti)-symmetrization procedure. At low orders in the intermolecular perturbation expansion, the \sapt contributions are given specific physical interpretations. For example, we may write the \sapt interaction energy as:
\begin{align} 
    \EintSAPT{} & = \sum_{j=0} \left[ \EELEC{1}{j} + \EEXCH{1}{j} \right] \nonumber \\
        & + \, \sum_{j=0} \left[ \EINDD{2}{j} + \EEXCHIND{2}{j} \right] + \dHF{2}  \nonumber \\ 
        & + \, \sum_{j=0} \left[ \EDISPP{2}{j} + \EEXCHDISP{2}{j}  \right],
    \label{eq:SAPT-components}
\end{align}
where the upper limits of the sums ($j$) will vary depending on the level of intramolecular correlation included. 
These terms are commonly regrouped according their physical meaning: electrostatic ($\{ \EELEC{1}{j} \}$), exchange-repulsion ($ \{ \EEXCH{1}{j} \} $), induction ($ \{ \EINDD{2}{j}, \EEXCHIND{2}{j} \} $) and dispersion ($ \{ \EDISPP{2}{j}, \EEXCHDISP{2}{j} \} $). In this illustration, the intermolecular perturbation expansion is conducted only to second-order. Higher-order effects are often important and are approximated using the delta-Hartree--Fock term, \dHF{n}, which approximates polarization and charge-delocalization effects from orders higher than included in the perturbation theory. 
Here $n$ is the maximum order of terms included in pure \sapt energies, so \dHF{n} will include contributions from order $n+1$ and higher. Since the \dHF{n} term is non-perturbative, thus it is not strictly a \sapt term. But it often represents a non-negligible contribution to the interaction energy such as strongly hydrogen-bonded complexes. It is commonly included as part of the total induction term. The \dHF{2} and \dHF{3} terms, respectively at second and third-order, are computed as: \cite{Jeziorska1987,moszynski_symmetry-adapted_1996,MasSBJ97} 
\begin{equation} 
    \dHF{2}  = \EintHF - (\EELEC{1}{0} + \EEXCH{1}{0} +           \EINDD{2}{0} + \EEXCHIND{2}{0})
    \label{eq:dhf2}
\end{equation}
and
\begin{equation} 
    \dHF{3}  = \dHF{2} - (\EINDD{3}{0} + \EEXCHIND{3}{0}).
    \label{eq:dhf3}
\end{equation}
Here \EintHF is the Hartree--Fock supermolecular interaction energy for the complex. The subscript ``r'' indicates that the response of interacting orbitals of each dimer is included in the induction terms (orbital relaxation effects) \cite{salter1987theory,trucks1988theory,trucks1988analytic,salter1989analytic}. 
Also, depending on the maximum order of the intra and inter-molecular perturbation used, we may define various levels of \sapt: \SAPT{0}, \SAPT{2}, \SAPT{2+}, \SAPT{2+(3)} and \SAPT{2+3} methods. The terms included in each of the \sapt levels are listed in \tabrff{tabsapt}.


\include{table-saptorder} 

On the other hand, \saptdft formulates the interaction energy contributions in terms of the density, density-response functions and interaction density matrices, all constructed from Kohn--Sham orbitals and orbital energies, with appropriate response kernels used for the density response functions. The advantage of this approach over the Hartree--Fock-based \sapt is both simplicity and accuracy. 
The \saptdft intermolecular interaction energy is the result of a single perturbation theory as the use of Kohn--Sham orbitals mitigates the need for the inclusion of intramolecular correlation effects. Consequently, the \saptdft interaction energy is written as:
\begin{align} 
    \EintSAPTDFT &= \Eelst + \Eexch \nonumber \nonumber \\
                 &~~~ \Eindpol{2} + \Eexind{2} + \dHF{2} \nonumber \\
                 &~~~ \Edisppol{2} + \Eexdisp{2}.
    \label{eq:Eint-SAPTDFT}
\end{align}
The numbers in the superscripts correspond to the order of intermolecular perturbation. 
The second-order energies are computed using coupled Kohn--Sham (CKS) response kernels. However, the one exception is \Eexdisp{2} which in \CamCASP is estimated from the uncoupled-CKS energy $\Eexdisp{2}[\mathrm{UCKS}]$ by scaling as:
\begin{align}
    \Eexdisp{2} &\approx \Eexdisp{2}[\mathrm{UCKS}] \times 
    \frac{\Edisppol{2}}{\Edisppol{2}[\mathrm{UCKS}]}, 
\end{align}
where $\Edisppol{2}[\mathrm{UCKS}]$ is (non-exchange) dispersion energy computed with the uncoupled-CKS kernel. Note that \Eexdisp{2} can also be computed without scaling. \cite{Jansen2014}


Due to difficulties in deriving the exchange terms of second and higher-orders in the perturbation operator, until very recently, these terms were computed in the \Ssq, or single exchange approximation (SEA). This approximation is known to breakdown when the wavefunction overlap of the interacting species increases. In this case, the \Ssq approximation results in too little exchange repulsion, and an unphysical overstabilization of the complex. This has been shown to get worse at higher-orders in perturbation theory.\cite{JeziorskiBP76a,SchafferJ12} 
As an empirical approach to partly alleviate the problem \cite{patkowski2007}, the higher-order exchange energies can be scaled by the multiplicative factor $\mathrm{p_{ex}}(\alpha)$ as: 
\begin{equation}
    \mathrm{p_{ex}}(\alpha) = \left(
       \frac{\EEXCH{1}{0}}{\EEXCH{1}{0}(\Ssq)}
                          \right)^{\alpha},
     \label{eq:parker-exch-scaling}
\end{equation}
where $\EEXCH{1}{0}(\Ssq)$ and $\EEXCH{1}{0}$ are the first-order exchange energies computed with and without the \Ssq approximation using the expressions from Jeziorski \etal \cite{JeziorskiBP76a}, and $\mathrm{p_{ex}}(\alpha)$ is a scale factor that can be modulated by the $\alpha$ exponent. The default choice $\alpha = 1$ is used for \sapt calculations as recommended by Parker \etal \cite{Parker2014}.

In this work, for the \saptdft \Eexind{2} energy in \eqrff{eq:Eint-SAPTDFT}, we have used a formulation of the theory in which the second-order exchange-induction energy is computed without the single-exchange approximation (\saptdftnossq). To do so, we have implemented in the \CamCASP code the spin-summed (closed shell) form of the expression derived by Schaffer \& Jansen. This has major consequences for the very strongly bound complexes we have investigated (see \secrff{sec:results}).
While the \Ssq approximation is still in use in our calculations of the second-order exchange-dispersion energy, this term is usually small enough \cite{SchafferJ_single-determinant-based_2013} that the simple scaling expression \cite{Misquitta2003} should be appropriate.





\subsection{Charge-delocalization energy in \sapt \& \saptdft}
\label{sec:CD_SAPT_SAPTDFT}

In \sapt and \saptdft, the interaction energy is defined as the sum of physically meaningful quantities, however these theories have nothing to say about the charge-delocalization energy.
Rather, the charge-delocalization and polarization energies are both part of the induction energy computed from \sapt/\saptdft, and some scheme must be used to separate these.
Therefore, Stone and Misquitta (`SM09') have proposed a first definition of the $n^{\rm th}$ order charge-delocalization energy \cite{Stone2009} as:
\begin{align} \label{eq:CD-SM09}
    \EcdSM{n} &= \EIND{n}[{\rm DC}] - \EIND{n}[{\rm MC}],
\end{align}
where $\EIND{n}[{\rm DC}]$ is the induction energy computed in the dimer centered (DC) basis, while $\EIND{n}[{\rm MC}]$ is the energy computed in the monomer centered (MC) basis.
Note that $n=2$ for \SAPT{0}, \SAPT{2}, \SAPT{2+}, and $n=3$ for \SAPT{2+(3)} and \SAPT{2+3}.
The idea here is that the dimer-centered basis 
allows the description of charge-delocalization-type excitations, while the monomer-centered basis does not. 
This is the CD definition used in the \PsiF package. But as discussed by Stone \& Misquitta, and demonstrated by Misquitta \cite{Misquitta13}, this definition has serious deficiencies for large basis sets and short intermolecular separations. In both cases, the monomer-centered basis sets can also describe CD-type excitations, thus leading to ever diminishing allocations of the induction energy to CD upon increasing the basis set. 

As an alternative, Misquitta has proposed \cite{Misquitta13} a regularization of the electrostatic potential as a means of defining the charge-delocalization. The induction energy is the response of a molecule to the electrostatic potential of the partner (or environment). This potential consists of a well-behaved, repulsive contribution from the electronic density and a singular, attractive contribution from the point nuclear charges. 
In this viewpoint, charge-delocalization is associated with electron tunneling into the singular nuclear potential, and hence can be suppressed by suitably eliminating the singularity in this potential. 
This can be done by using a Gaussian screening function to split the $1/r$ nuclear potential into a singular and regularized part \cite{PatkowskiJS01a} such as: 
\begin{align}
  \frac{1}{r} = v_p(r) + v_t(r),
  \label{eq:Vreg}
\end{align}
where $v_t = \frac{1}{r}\left( 1 - e^{-\eta r^2} \right)$ is the singular, short-ranged part, and $v_p = \frac{1}{r} e^{-\eta r^2}$ the long-ranged, well-behaved part of the nuclear potential. 
The regularized induction energy is then computed using the well-behaved, regularized nuclear potential. 
With a suitable choice of the parameter $\eta$, it has been shown that all of the charge-delocalization can be suppressed leading to a `pure' polarization energy. Hence, we may define the charge-delocalization energy at order $n$ as: 
\begin{align}
  \EcdREG{n} &= \EIND{n} - \EINDreg{n}.
  \label{eq:CD-Reg}
\end{align}
Note that Misquitta has determined that $\eta=3.0$ a.u.\ is a suitable choice for a range of molecular systems, though we may expect this parameter to vary with system, albeit to a small extent. The possible dependencies of $\eta$ upon the nature of the interacting partner(s) will be studied in a separate work.

 
\subsection{Polarization models and higher-order charge-delocalization \\ energies within \saptdft}
\label{sec:higher-order_CD}

Presently, using the SM09 method, the charge-delocalization energy from \sapt can be computed at second and third-order only. And if computed using regularized \saptdft, this can be done to second-order only.
This poses a problem since there are contributions to induction from higher-order terms, and these can be as important as the second or third-order induction terms. 
Such higher-order contributions are often estimated using the \DHF correction. However, as this energy correction is computed in a hybrid approach that combines low-order \SAPT{0} with supermolecular Hartree--Fock, there is at present no way to decompose the \DHF term into separate polarization and charge-delocalization. It is actually not clear that such a decomposition is even theoretically feasible, as these effects are sure to couple at higher-orders in perturbation theory.
Nevertheless, because of the relative size of the \DHF correction (it is nearly as large as the second-order induction for the water dimer at equilibrium), the charge-delocalization component of this term must be included if we are not to severely underestimate the charge-delocalization from the \sapt or \saptdft approaches.

Recently a method to extract the charge-delocalization component from the \DHF energy correction was proposed by Misquitta \& Stone \cite{MisquittaS16} and used for the pyridine dimer. In parallel with this work, Gilmore, Stone \& Misquitta \cite{GilmoreDM_water_2020} have used this method on the water dimer.
We describe below the relevant details of the polarization models used in this paper. 

The classical polarization energy of an ensemble of molecules (or units) is $\EPOLCL = \sum_A \EPOLCL(A)$, where $A$ is the molecular label. The classical polarization energy of a molecule $A$ is defined as:
\begin{align}
  \EPOLCL(A) &= \frac{1}{2} 
               \sum_{a \in A} \sum_{B\ne A} \sum_{b \in B} \sum_{tu}
                \Delta \Q{a}{t} f_{n(tu)}(\BETApol{ab} R_{ab}) \T{ab}{tu} \Q{b}{u},
                \label{eq:pol_classical}
\end{align}
where 
$a$ ($b$) labels sites in molecule $A$ ($B$),
the ranks of the moments are given in the compact form  $t \equiv l\kappa$ where $l=0,1,2,\dots$ is the angular momentum quantum number, and $\kappa=0,1c,1s,\dots,lc,ls$ labels the real components of the spherical harmonics of rank $l$ (see Appendix B in Stone \cite{Stone:book:13}).
$\Q{a}{t}$ is the multipole moment operator for moment $t$ at site $a$, 
and $\T{ab}{tu}$ is the interaction tensor \cite{Stone:book:13} which describes the interaction between a multipole $\Q{b}{u}$ at site $b$ and a multipole $\Q{a}{t}$ at site $a$.
$f_{n(tu)}(\BETApol{ab} R_{ab})$ is a damping function of order $n$ defined below in \eqrff{eq:TT}, where $\BETApol{ab}$ is the damping parameter for the $(ab)$ site pair, and $R_{ab}$ is the distance between these sites.
Here $n$ is a function of the tensor ranks $t$ and $u$, and if $t=l_1\kappa_1$ and $u=l_2\kappa_2$, then $n=l_1 + l_2 + 1$.
Here, we assume that the damping function depends only on the distance $R_{ab}$ between the sites  and not on their relative orientation.
$\Delta \Q{a}{t}$ is the change in multipole moment $t$
at $a$ due to the self-consistent polarization of site $a$ in the field of all sites on {\em the other} molecules such as:
\begin{align}
  \Delta \Q{a}{t} &= - \sum_{a' \in A} \sum_{B\ne A} \sum_{b \in B} \sum_{t'v}
                   \A{aa'}{tt'} f_{n(t'v)}(\BETApol{a'b} R_{a'b})
                   \T{a'b}{t'v} (\Q{b}{v} + \Delta \Q{b}{v}),
                \label{eq:deltaQ}
\end{align}
where $\A{aa'}{tt'}$ is the distributed polarizability for sites
$(a,a')$ which describes the response of the multipole moment component $\Q{a}{t}$ at site $a$ to the $t'$-component of the field at site $a'$.
To find $\Delta \Q{a}{t}$ we need to solve \eqrff{eq:deltaQ} iteratively.
If $\Delta \Q{b}{v}$ is dropped from the right-hand-side of this equation then the resulting $\Delta \Q{a}{t}$, when inserted in \eqrff{eq:pol_classical} leads to the second-order polarization energy, \EpolCL{2}.
In the models constructed here, we have assumed the localized forms of the distributed polarizabilities: $\A{aa'}{tu} = \A{a}{tu} \delta_{aa'}$. The localization is performed using the techniques described by Stone \& Misquitta \cite{misquitta_isa-pol_2018,MisquittaS08a}.
For the polarization damping we have used the Tang--Toennies \cite{TangT84} damping model:  
\begin{align}       
  f_n(x) &= 1 - e^{-x} \sum_{k=0}^{n} \frac{x^k}{k!},
  \label{eq:TT}                       
\end{align}
where the parameter $x \equiv x_{ab}$ is either related to the site-site separation $R_{ab}$ linearly as $x_{ab} = \BETApol{ab} R_{ab}$, or we use the Slater functional form (SlaterFF) relationship \cite{VanVleetMSS15}:
\begin{align}
    x_{ab} &= \BETApol{ab} R_{ab} -
      \frac{\BETApol{ab} R_{ab}(2\BETApol{ab} R_{ab} + 3)}
        {(\BETApol{ab} R_{ab})^2 + 3 \BETApol{ab} R_{ab} + 3}.
        \label{eq:slaterFF}
\end{align}
The SlaterFF relationship has the effect of introducing a separation-dependent damping coefficient and can be beneficial for complexes at very short separations.
There could well be other forms for $x_{ab}$, but these choices were sufficient for the complexes studied in this work.

As we derive the distributed multipoles and distributed anisotropic (localized) polarizabilities using the BS-ISA (Basis-Space Iterated Stockholder Atoms) \cite{misquitta_distributed_2014-1} and ISA-Pol \cite{misquitta_isa-pol_2018} algorithms, the only unknowns in the above definition of the classical polarization energy are the damping parameters. These are determined by fitting the non-iterated classical polarization energy \EpolCL{2} to a selection of regularized second-order induction energies \EINDreg{2}. 
Subsequently, the infinite-order classical polarization energy, \EpolCL{2-\infty}, is determined by iterating the classical polarization model to convergence.
If we approximate the infinite-order induction energy from \saptdft as:
\begin{align}
    \EIND{2-\infty} \approx  \EIND{2} + \dHF{2},
      \label{eq:Eind-infinite-order}
\end{align}
then the infinite-order charge-delocalization energy is defined as:
\begin{align}
    \Ecd{2-\infty} &= 
        \EIND{2-\infty} - \Epol{2-\infty} \nonumber \\
        &\approx \EIND{2} + \dHF{2} - \EpolCL{2-\infty}.
    \label{eq:CD-infinite-order}
\end{align}
This approach results in a definition of the charge-delocalization energy that is dependent on the polarization model, but from our experience, this dependence is relatively small in practice. From now on, we denote \saptdft/Pol-Model, the energies obtained using this approach.


\section{Results and Discussion}
\label{sec:results}

\subsection{Accuracy of $\EINT$}
It is instructive to assess the accuracies of \EINT from supermolecular EDA and from \sapt methods, in their present stage of development, as compared to \ccsdt results. The present study focuses on the interactions of strongly bound complexes, in which polarization and/or charge-delocalization are expected to contribute significantly to \EINT.
\subsubsection{Comparison of SAPT, SAPT(DFT) and DFT}
\label{sec:accuracy_of_Eint}
\include{Eint_water_zn} 
In \figrff{fig:Zn-H2O-Eint}, we represent the distance evolution of the intermolecular interaction energies of three complexes of a water molecule: with \Zn, with \F and with another \water molecule. Similar plots for a water molecule with \Cl and \OH are in Figure S2 in the \SI. It is first observed that the three hybrid functionals, \pbez, \wbxd and \blyp, can lead to significant differences in \EINT. For \waterdimer, \pbez and \wbxd are nearly identical, even though \pbez lacks the dispersion correction, and both methods are in good agreement with the \ccsdt references at longer separations. As the long-range energy of water is dominated by the electrostatic interaction, this is evidence that the densities from these methods are quite accurate. However, \blyp tends to underestimate the interaction energy at all separations (\subfigrff{fig:Zn-H2O-Eint}{A}).
For \Fwater, all three density functionals give accurate interaction energies. \wbxd and \blyp are the most accurate, providing virtually identical interaction energies, while \pbez overestimates the interaction energy at shorter separations (\subfigrff{fig:Zn-H2O-Eint}{E}). 
For \Znwater, all functionals overestimate the interaction energy, with \wbxd showing the smallest errors (around $3$\%), followed by \pbez ($5$\%), and \blyp ($7$\%). At the largest separation, \wbxd shows closest agreement with \ccsdt, but both \pbez and \blyp diverge from \ccsdt where \EINT being more negative and a local maximum appears around 3.5 \AA. Further convergence was also not possible with all \dft functionals for some geometries at these larger separations (\subfigrff{fig:Zn-H2O-Eint}{C}).
An investigation of the charge on the zinc cation shows that the excessive charge delocalization is responsible for these issues.
The charge on \Zn has been computed using an Iterated Stockholder Atoms (ISA) analysis with the \CamCASP code, and AIM (Atom in Molecule) analysis with the \Gaussian code. 
We observe that with a Hartree--Fock wavefunction the charge on \Zn is $+2e$ at long-range, but gets smaller as separation decreases. For example, at $3.0$ \AA, the charge on \Zn is $+1.95e$, but with the density functionals this charge is $+1.76e$ (\blyp), $+1.79e$ (\pbez), and $+1.84e$ (\wbxd). These are all smaller than the charge from HF, with the largest deviations from \blyp and \pbez. Further, as separation increases, the charge on \Zn still decreases with the density functionals. For example, by $4.6$ \AA\xspace the zinc charge from \wbxd decreases to $1.79e$ while the HF charge was already nearly $+2e$ by $3.5$ \AA. This means that all density functionals result in an increased, spurious electrostatic interaction between the \Zn cation and the negatively charged \water molecule. This is seen in single-determinant HF and so is not a multi-reference issue, but is more likely related to the delocalization error in \dft. As the \wbxd functional is a range-separated functional it is more immune to this error, but nevertheless still shows some signs of spurious charge delocalization.

On the other hand, all \sapt methods show high accuracies for all of these systems, including the \Znwater complex. In all cases, \sapt methods exhibit the correct long-range form, with errors showing up only at very short separations. Indeed,  \SAPT{2+3} shows over-binding compared to \SAPT{2} for all three complexes (\subfigrff{fig:Zn-H2O-Eint}{B, D, F}). This was not expected as \SAPT{2+3} includes terms of third order in the intermolecular perturbation operator, but there are theoretical reasons for its lower accuracy (see \secrff{sec:effect_of_S2_approx_on_Eint}). 
Interaction energies from \saptdftnossq are in consistently good agreement with the reference \ccsdt energies. All \saptdft results were performed using asymptotically corrected \pbez. We have demonstrated in the \SI, that contrary to previous statements in the literature \cite{Lao2015} this correction is theoretically needed regardless of the charge state of the interacting species (see Table S1 and Figure S3). 
The only systematic weakness of \saptdft and other \sapt methods is the over-binding at very short separations of $1$ \AA \xspace in the anions $\dots$ \water interactions (\subfigrff{fig:Zn-H2O-Eint}{F} and S2).

The overall excellent performance of \saptdft and also \SAPT{2} make these theories pass the first test of applicability for \polFF development. They are consistently accurate for the strongly polar systems of interest, and unlike \pbez, \blyp and \wbxd, they can be used on all systems (anionic, cationic and neutral) with no strong systematic errors (see also Figure S2 for supporting data on \Clwater and \OHwater complexes).

\input{SAPT_SAPTdft_and_dHF_S2_and_noS2}  

\subsubsection{Effect of the \Ssq approximation on \EINT and \DHF}
\label{sec:effect_of_S2_approx_on_Eint}
In this and subsequent sections we examine the implications of the \Ssq approximation on the interaction energies and energy components. As mentioned in \secrff{sec:SAPT}, Sch\"{a}ffer \& Jansen have shown that the \Ssq approximation causes the second-order exchange energies to be systematically underestimated. And this error gets worse at shorter bond lengths for which orbital overlap effects are larger. The \Ssq approximation also has an effect on the \DELTAHF{n} correction (see \eqrff{eq:dhf2} and \eqrff{eq:dhf3}). \figrff{fig:SAPT_SAPTDFT_and_dHF_S2_and_noS2} shows in yellow the difference between \Eint{n} (pure \sapt) and \Eint{n}+\dHF{n} for \SAPT{2+3} and \saptdft interaction energies. First, we consider the \SAPT{2+3} results. For \waterdimer, \Eint{3} is already close to the \ccsdt references, but upon including \DELTAHF{3}, the total $\Eint{3}+\DELTAHF{3}$, now over-binds, with the location of the minimum and the repulsive wall both moving to shorter O \dots H separations (\subfigrff{fig:SAPT_SAPTDFT_and_dHF_S2_and_noS2}{A}).
For \Znwater (\subfigrff{fig:SAPT_SAPTDFT_and_dHF_S2_and_noS2}{C}) and \Fwater (\subfigrff{fig:SAPT_SAPTDFT_and_dHF_S2_and_noS2}{E}), the pure \sapt energy \Eint{3} shows no minimum, but here the apparent short-range divergence to negative energies is even larger. For neither of these systems does \Eint{3} show a minimum. Rather, for \Znwater, \Eint{3} appears to diverge, with an energy below $-500$ \kcalmol at the shortest separation. For \Fwater, \Eint{3} gets close to $-300$ \kcalmol at the shortest separation. The \DELTAHF{3} correction, which is predominantly positive for both systems, once again ``fixes'' the large over-binding of the \Eint{3} energies. This results in good agreement with \ccsdt for \Znwater, but for \Fwater $\Eint{3}+\DELTAHF{3}$ from \SAPT{2+3} still over-binds by around 7\% at the minimum and more on the repulsive wall (see the data tables in the \SI).

With \saptdft, there is a significantly different outcome. For \waterdimer, both \Eint{2} and $\Eint{2}+\DELTAHF{2}$ show a minimum, the latter being close to the \ccsdt reference (\subfigrff{fig:SAPT_SAPTDFT_and_dHF_S2_and_noS2}{B}). 
For \Znwater, in contrast to \Eint{3}:\SAPT{2+3} energy, \Eint{2}:\saptdft shows a minimum reasonably close to that from \ccsdt (\subfigrff{fig:SAPT_SAPTDFT_and_dHF_S2_and_noS2}{D}), while it over-binds at short separations, this is by less than 10\% near the minimum, with no apparent divergence even at the shortest separation of $1.5$ \AA. However, upon inclusion of the \DELTAHF{2} term, the 10\% error near the minimum is reduced to less than $2$\%, and the $\Eint{2}+\DELTAHF{2}$:\saptdft energies are very close to the \ccsdt reference, with the equilibrium distances being nearly identical as well.
For \Fwater, \saptdft also results in well-behaved interaction energies at the pure \sapt level, \Eint{2}, and upon inclusion of \DELTAHF{2}, gives $\Eint{2}+\DELTAHF{2}$ energies in very good agreement with \ccsdt (\subfigrff{fig:SAPT_SAPTDFT_and_dHF_S2_and_noS2}{F}).
Going further, this analysis has also been conducted at the \SAPT{2} level for these three complexes. Briefly, we observe a similar behaviour as with \saptdft for \waterdimer and \Fwater, but for \Znwater, \SAPT{2} shows similar problems as made by \SAPT{2+3} (Figure S4).

Overall, some favorable features are worth noting. Firstly, the consistently good behaviour of \saptdft for these strongly bound systems is encouraging: the \saptdft interaction energies are not only reliable as a total, but this total is built from parts themselves well-behaved and thus does not rely on error cancellation. This is the case even at the very short, and nearly chemical bonding separations. 

Secondly, the good performance of \saptdft is to a large extent due to the removal of the \Ssq approximation in the major exchange terms. 
If the \Ssq approximation were present in the \Eexind{2} and \DELTAHF{2} energies, \saptdft would behave like \SAPT{2}. This can be seen from a comparison of Figure S4 and S5 in the \SI, where we have displayed \saptdft energies with the \Ssq approximation used in the second-order exchange-induction energies. This is encouraging as it implies that if the \Ssq approximation is removed then \SAPT{2} and also \SAPT{2+3} could be as well viable alternatives to \saptdft.

In Figures S6 and S7, we have illustrated the extent of the \Ssq error in the \EIND{2} energy, and also in the sum $\EIND{2}+\DELTAHF{2}$. 
However for all complexes, except at very short separations, the \Ssq error is nearly completely cancelled when \DELTAHF{2} is added to \EIND{2}, in agreement with the results of Sch\"{a}ffer \& Jansen.
Why then do we need bother with the \Ssq approximation? The answer is that if we are interested in the individual contributions to the interaction energy, then as demonstrated next, this approximation must be removed.
\subsection{Separability of the interaction energy}\label{sec:separability}

\include{Comparison_of_SAPT_SAPTDFT_and_EDAs} 

The contributions of the interaction energy from  SAPT-based methods and supermolecular EDA are compared below. Such comparisons should enable for an informed choice in \polFF development to parametrize each separate physical motivated contributions.

\subsubsection{Comparison of SAPT, SAPT(DFT) and EDAs}
\label{sec:comparison_of_SAPT_SAPTDFT_and_EDAs}


The sum of electrostatic and exchange-repulsion energies, $\ELST+\EXCH$, is first analyzed for three of the complexes (\figrff{fig:Comparison_of_SAPT_SAPTDFT_and_EDAs}).
We observe that all three \sapt methods give essentially the same energies at all separations, and the supermolecular EDAs are also in mutual agreement. However, there are small but appreciable differences between \sapt and supermolecular EDAs. In the latter, values are always more stabilizing than those from \sapt. For the \waterdimer, \Znwater, and \Fwater complexes at their equilibrium separations, the difference in $\Eelst+\Eexch$ from \saptdft and $\ELST+\EXCH$ from \almofunc{\wbxd} is slightly larger than $1$ \kcalmol (\subfigrff{fig:Comparison_of_SAPT_SAPTDFT_and_EDAs}{A}), $3$ \kcalmol (\subfigrff{fig:Comparison_of_SAPT_SAPTDFT_and_EDAs}{C}), and $5$ \kcalmol (\subfigrff{fig:Comparison_of_SAPT_SAPTDFT_and_EDAs}{E}), respectively. 
Such differences of the first-order energies between SAPT and supermolecular EDA methods could be caused by the fact that the latter actually compute the Heitler--London energy. At the HF level, the difference is of the order of $S^4$ and is almost negligible, but at the \dft level, it is of the order of $S^2$ and can be large \cite{podeszwa2010}.
Overall, these differences are not negligible when compared with \EINT, and may well have consequences for the development of \polFF with a physical representation of the interaction energy components.

We next consider the total induction energies (\subfigrff{fig:Comparison_of_SAPT_SAPTDFT_and_EDAs}{B, D, F}). For all complexes, the DFT-based supermolecular EDAs give induction energies that are systematically more negative than the SAPT-based ones. \almofunc{\pbez} and \csovblyp give nearly the same energies, while \almofunc{\wbxd} gives induction energies that are somewhat closer to those from the \sapt methods.
This is this likely due to the self-interaction error which results in an over-polarization of the system. As this error is known to decrease with increasing fraction of Hartree--Fock exchange, we should expect the induction energies (in magnitude) to be ordered as follows: \csovblyp > \almofunc{\pbez} > \almofunc{\wbxd} > \sapt, which is indeed the case.
In the case of \Znwater (\subfigrff{fig:Comparison_of_SAPT_SAPTDFT_and_EDAs}{D}), the three density functionals EDAs are not only off-set from the \sapt methods as explained above, but appear to diverge from each other. The DFT-based supermolecular EDAs give induction energies that are systematically more negative than the SAPT-based ones by between $7$ to $10$ \kcalmol for the range of \Zn \dots O distances. This difference does not get smaller with increasing separation, but actually {\em increases} with \almofunc{\pbez} and \csovblyp. This effect is unphysical and could pose a significant problem for building polarization models, and we will elaborate on this issue in the next section. In contrast, \rvshf gives induction energies that are smaller in magnitude than the \sapt values, and these seem to converge to the \sapt energies at long-range.
About SAPT-based methods: \SAPT{2}, which has identical total induction with \SAPT{2+3}, and \saptdft are in overall agreement, with small differences evident only at the shorter intermolecular separations (see also data tables in the \SI). This is no doubt due to the cancellation of errors from the \Ssq approximation in the pure \sapt energies and the \DHF terms in \SAPT{2} and \SAPT{2+3}. In general, \saptdft gives less negative induction energies than \sapt{2}, with the largest differences (at equilibrium separations) being $3.9$ \kcalmol for the \Fwater complex (\subfigrff{fig:Comparison_of_SAPT_SAPTDFT_and_EDAs}{F}).

Finally, we examine the trends of the dispersion contribution from the SAPT-based methods. For all three complexes (see Figure S8 in the \SI), the \EDISP{2} dispersion energies from \SAPT{2} and \saptdft are in good agreement. But the \SAPT{2+3} \sEDISP{3} values are consistently more negative than the \EDISP{2} ones. While this is not surprising, the unusual behaviour of \sEDISP{3} for the \Znwater complex (Fig. S7D) does give us cause for concern. Indeed, \sEDISP{3} is substantially more negative than \EDISP{2} from either \SAPT{2} or \saptdft. But for separations smaller than $1.9$ \AA\, it shows a minimum and then gets less negative than \EDISP{2}. This is unexpected and likely an unphysical consequence of the \Ssq approximation that has a much larger effect on the third-order exchange-dispersion and mixed induction-dispersion exchange energies that are included in \sEDISP{3} (see \tabrff{tabsapt}).

\input{CD_POL_from_SAPTDFT_and_EDAs}  
\subsubsection{\ECD \& \EPOL: variations with methods}
\label{sec:CD_variations_with_methods}

In this section, we address the separability of the induction energy into polarization (POL) and charge-delocalization (CD) in the \sapt and supermolecular EDAs methods shown in Figures 5 and 6.

We first compare POL and CD energies obtained with the widely used Stone \& Misquitta (SM09) method (\SAPT{2}, \SAPT{2+3}), and the newer regularized charge-delocalization definition (\saptdft). 
We observe that the POL and CD energies from \SAPT{2} and \SAPT{2+3} are effected by both the \Ssq approximation as well as by the use of the SM09 algorithm. As explained above, the \Ssq approximation leads to an underestimation of the exchange-induction energies, and consequently increases POL and CD (in magnitude). The SM09 basis-space algorithm for defining the POL and CD energies can be ill-defined when large basis sets are used, or when the complex separation is small. In both cases, it has been shown \cite{Misquitta13} that SM09 will lead to more POL and less CD (in magnitude). 
Therefore, the POL energies from \SAPT{2} and \SAPT{2+3} using both the \Ssq approximation and SM09 are more negative (\subfigrff{fig:POL_from_SAPTDFT_and_EDAs}{A, C, E}), but the CD energies are likely less due to a cancellation of errors (\subfigrff{fig:CD_from_SAPTDFT_and_EDAs}{A, C, E}).
On the other hand, in \saptdft, the increased exchange energy present when the \Ssq approximation is removed causes \Ecd{2}/no\Ssq energies to decrease for all systems as expected. But the removal of the \Ssq approximation has relatively little effect on \Epol{2}/no\Ssq except at the shortest separations for \Znwater and \Fwater. In both complexes, \Epol{2}/no\Ssq energies are larger in magnitude than \Epol{2}/\Ssq (\subfigrff{fig:POL_from_SAPTDFT_and_EDAs}{C, E}). And \Epol{2}/\Ssq shows an unexpected increase at very short-range (in magnitude) which is not observed in the \Epol{2}/\noSsq energies. We do not as yet understand why this is the case. 
We suspect that it is an artifact of the regularization potential, which, at these very short separations, is likely to suppress the polarization energy. However, this does not explain why this feature is not present in \Epol{2}/no\Ssq.
Overall, \saptdft/\noSsq gives consistently physically acceptable estimates for \Epol{2} and \Ecd{2} for all three complexes at all intermolecular separations.

Next, since the POL and CD energies from supermolecular EDAs are computed non-perturbatively, these energies are effectively at infinite-order in the context of intermolecular perturbation. Therefore, when making comparisons with \saptdft, it is essential to ensure that the infinite-order CD and POL energies are used to avoid misleading assessments.\cite{Misquitta13,MisquittaS16,GilmoreDM_water_2020,Mao2018} 
We have done this using the methodology described in \secrff{sec:higher-order_CD}, through which the infinite-order energies are defined with the help of classical polarization models. Details of the construction of these models and their parameters can be found in the \SI (Figures S9 and S10). The infinite-order polarization (\Epol{2-\infty}) and charge-delocalization (\Ecd{2-\infty}) energies have then been estimated for \saptdft/no\Ssq.
In all systems, the inclusion of higher-order effects in the \saptdft POL (\subfigrff{fig:POL_from_SAPTDFT_and_EDAs}{B, D, F}) and CD (\subfigrff{fig:CD_from_SAPTDFT_and_EDAs}{B, D, F}) energies brings them closer to the corresponding energies from the supermolecular EDAs. Differences nevertheless remain since the \saptdft POL and CD energies are smaller in magnitude than those from the \dft-based supermolecular EDAs, and often in good agreement with the Hartree--Fock based RVS EDA.
For the water dimer, \Ecd{2-\infty} from \saptdft agrees well with the \rvshf result, but has $0.5$ to $1$ \kcalmol lesser magnitudes than the \dft-based EDA methods at the energetically relevant dimer separations (\subfigrff{fig:CD_from_SAPTDFT_and_EDAs}{B}). 
For \Fwater, there is a much closer agreement between \saptdft and all supermolecular EDAs regarding \EPOL (\subfigrff{fig:POL_from_SAPTDFT_and_EDAs}{F}). However, as with the water dimer, there are large differences (from $2$ to $4$ \kcalmol) between the CD energies from \saptdft and the three DFT-based methods (\subfigrff{fig:CD_from_SAPTDFT_and_EDAs}{F}).
Though these differences are significant, they are much smaller than the differences between the second-order POL and CD energies from \saptdft and the corresponding energies from the supermolecular EDAs. Therefore, we do see a convergence of these methods, but only when
the \Ssq approximation is removed, the CD/POL split is defined through the use of regularization, and infinite-order effects are included.

Finally, we discuss the significant variations of POL and CD energies in the \sapt and supermolecular EDA methods seen in the \Znwater complex. First, we observe a significant overestimation of the polarization energies from \SAPT{2} and \SAPT{2+3} when compared to the other methods (\subfigrff{fig:POL_from_SAPTDFT_and_EDAs}{C, D}). And the CD energies from \SAPT{2} are close to zero except at very short separations (\subfigrff{fig:CD_from_SAPTDFT_and_EDAs}{C}). This arises from the competing \Ssq and basis-set effects of the SM09 approach as mentioned above. 
Second, we see larger differences between \saptdft and the DFT-based supermolecular EDAs. It was previously observed, for $\mathrm{R_{\Zn \dots O}} < 2.5$ \AA, that both \almo and \csov result in total induction energies about $7$ to $10$ \kcalmol more negative than \sapt (\subfigrff{fig:Comparison_of_SAPT_SAPTDFT_and_EDAs}{D}). This difference originates mainly from the polarization energy which is about $10$ \kcalmol more negative than with the DFT-based supermolecular EDAs. However, this difference gets smaller upon inclusion of infinite-order effects in \saptdft (\subfigrff{fig:POL_from_SAPTDFT_and_EDAs}{D}).
Lastly, the \Znwater complex shows unusual features in the CD energy (\subfigrff{fig:CD_from_SAPTDFT_and_EDAs}{D}). In the context of \saptdft, \Ecd{2} and \Ecd{2-\infty} have different radial dependencies, with only \Ecd{2} showing a pure exponential decay. Nevertheless, \Ecd{2-\infty} does decrease in magnitude with distance while this is not the case for the three DFT-based supermolecular EDAs. The values of CD from \csovblyp, \almofunc{\pbez} and \almofunc{\wbxd} actually increase in magnitude with distance. This is unphysical and is related to the excessive charge delocalization of the \Zn cation in the \Znwater complex in \dft as discussed in \secrff{sec:comparison_of_SAPT_SAPTDFT_and_EDAs}.



The unusual behaviour of the CD energy for $\mathrm{R_{\Zn \dots O}} < 2.5$ \AA \xspace needs to be commented on: The CD energy from the \dft and HF methods show a non-exponential form, with a maximum just below 2.0 \AA. 
This feature is more clearly visible in Figure S9 which shows that it is present in the HF energy no matter which EDA is used. 
As it is present in both HF and \dft, it cannot be because of the well-known self-interaction error in the Kohn--Sham formalism.
In fact, this apparently un-physical behaviour of the CD energy was previously observed in early Hartree-Fock \rvs studies on complexes of divalent cations with anionic ligands \cite{gresh1995energetics,gresh1996comparative}, and more recently for actinide complexes \cite{Gourlaouen2013,MARJOLIN201325}. The present supermolecular EDA interaction energy curves are not the diabatic interaction energy curves that one would expect. It has been shown in many systems involving metal (Mn$^{2+}$ or Mn$^{3+}$) cations and water that the system ionized to Mn$^+$ (or Mn$^{2+}$) and \water$^+$ \cite{Corongiu1978,Gourlaouen2013,MARJOLIN201325}. This effect is observed at the Hartree--Fock level but is prevented in pure \sapt as the formalism enforces closed-shell systems with fixed charged states. However, it re-enters the \sapt energy through the \DHF energy, and so the infinite-order CD energies from \saptdft also display a non-exponential behaviour at short-range.
In practice, \dft results tend to exhibit an even larger deviation from exponential decay due to spurious self-interactions \cite{Mori-Sanchez2006,LucasBao2018}.

Overall, \saptdft estimates for POL and CD energies have in most cases smaller magnitude than either the \almo or the \csov ones. Such differences are not sensitive to the choice of functional, whether hybrid or range-separated. But one needs to take into account such issues encountered at short and long-range of separations when developing \polFF. 
\section{Conclusions and perspectives}
We have performed joint \sapt and supermolecular EDAs calculations on a set of strongly bound and challenging complexes: \waterdimer, \Znwater and \Fwater, with some additional data provided for the \Clwater and \OHwater complexes. We have unravelled the similarities and differences between these methods in order to define an \abinitio framework in order to enable the  calibration and validation of accurate and separable polarizable force fields. We have paid particular attention to the separation of the induction energy into polarization and charge-delocalization as this is still a contentious issue. In order to address it, we have determined the infinite-order polarization and charge-delocalization energies using \saptdft, \regsaptdft and classical polarization models developed for the three main complexes studied here.
We have demonstrated that the use of \regsaptdft is to be preferred over the older basis-space algorithm from Stone \& Misquitta \cite{Stone2009}, particularly for strongly bound systems and at short intermolecular separations.
Comparisons with the supermolecular EDAs show that the  \Epol{2-\infty} and \Ecd{2-\infty} energies from \saptdft are much closer to the \almo and \csov ones than are the second-order energies. This is gratifying as these results strongly suggest that there is a convergence of concepts.

Moreover, following the work of Sch{\"a}ffer \& Jansen, it is now possible to compute the exchange-induction energy at second-order without the \Ssq approximation, and this has been implemented within \saptdft in the \CamCASP code. This approximation has been known to be invalid at short separations. In this connection, we have observed that the \Ssq approximation leads to an unusually strong underestimation of the exchange-induction for the very strongly bound systems we have studied. 
At third-order, the \Ssq approximation can be even more detrimental. At the time this work was conducted, there were no third-order exchange expressions free of the \Ssq approximation, however, recently Waldrop and Patkowski have recently derived third-order exchange-induction without this approximation \cite{Waldrop2021}, consequently it is now possible to compute the most important exchange terms in both \SAPT{2} and \SAPT{2+3} without using this approximation. Doing so would put these SAPT methods on par with \saptdft for strongly bound complexes. 

Regarding the supermolecular EDAs: Strong similarities were demonstrated between \csovblyp, \almofunc{\wbxd}, and \almofunc{\pbez}. Even with different wavefunctions, these EDAs are in quite good agreement showing similar results for the sum of electrostatic and exchange-repulsion, polarization and charge delocalization. However, there are small, but significant differences between \saptdft interaction energy components and those from the supermolecular DFT-based EDAs. These differences appear in the sum of the electrostatics and first-order exchange repulsion energies, with all EDAs leading to more negative energies compared with the \saptdft ones. 
The total induction energies from these supermolecular EDAs are larger in magnitude compared to \saptdft, the largest differences occurring for the \Znwater complex. But overall, energies from the EDAs and \saptdft agree asymptotically, except for the CD energies in \Znwater.

Finally, based on the complexes studies here, our recommended \abinitio references to develop \polFF based on the separability of \EINT into well-defined physical components are the \saptdftnossq and \almofunc{\wbxd} methods. Firstly, \saptdftnossq has been demonstrated to satisfy all requirements, even at the very short intermolecular separations of $1$ \AA\ or less. Contrary to what one might {\it a priori} suppose, the SAPT formalism does give meaningful and accurate results even at these short separation. 
Secondly, of the supermolecular EDAs, \almofunc{\wbxd} shows the highest accuracy and the most reliable partitioning. We found this method to be comparable to \saptdftnossq in accuracy, with interaction energy components from the two techniques in broad agreement, with the exception of the important \Znwater complex.
Lastly, the results of this paper led us to propose some recommendations for \saptdft calculations in the form of a check-list for force fields developers which are provided in the \SI.

\section{Technical Appendix}

\textbf{Numerical details.} All calculations have been performed using the aug-cc-pVTZ basis set for the water molecule, the zinc cation and anions (fluoride, chloride and hydroxy). \ccsdt full electrons calculations were performed with \Gaussian (D01 version) \cite{frisch2009g09} using the counterpoise method to correct the BSSE in the total intermolecular interaction energy. 
\rvs, \almo and \csov calculations were performed with the \Gamess, \cite{marquez1999quantum} \Qchem \cite{Shao2015} and \Hondo \cite{HONDO,Piquemal2005} softwares, respectively. \SAPT{2} and \SAPT{2+3} calculations were performed using the \PsiF package (version 1.1) and \saptdft with the \CamCASP program (version 6.0). Both SAPT calculations were carried out within the dimer centered basis. \saptdft calculations were performed using the hybrid ALDA+CHF (Adiabatic Local Density Approximation with coupled Hartree--Fock) kernel with orbitals and orbital energies from the \pbez \cite{perdew1996,AdamoB99a}, and with the Casida--Salahub \cite{Casida2000} asymptotic correction (AC). Previous studies \cite{Misquitta2002,Misquitta2005,Hebelmann2002,Hebelmann2003} have shown good results with the \pbez functional. It was therefore used for \saptdft calculations. Similarly, the \wbxd functional \cite{chai2008long} was used in \almo calculations. 
Basis sets and other numerical settings used for the calculation of distributed multipoles and distributed polarizabilities using the BS-ISA \cite{misquitta_distributed_2014-1} and ISA-Pol \cite{misquitta_isa-pol_2018} algorithms with the \CamCASP code are provided in the \SI.
Due to limitations of the implementation in the \Hondo code, the \blyp functional\cite{stephens1994} was used for \csov calculations, while with ALMO we will additionally use \pbez. Comparisons between the two methods will therefore not be completely consistent, but as both use hybrid density functionals, their results may be expected to be comparable. 
Also, only the polarization and charge-delocalization are reported for \csov calculations, since \rvs and \csov give the same values for electrostatic and exchange-repulsion.

\input{table-notation} 

\begin{suppinfo}
Supporting information contains:
(1) Additional data for the \Clwater and \OHwater complexes, and an analysis of the asymptotic correction for anions, 
(2) Additional data on the role of the \Ssq approximation,
(3) Additional data on the separability of the interaction energy,
(4) Details of the polarization models used in this paper,
(5) Additional figures illustrating the sensitivity of \saptdft to the choice of wavefunction, 
(6) A list of recommendations for \saptdft calculations of the interaction energy, and
(7) Energy tables and the complex geometries.
\end{suppinfo}
\section*{Acknowledgements}
This work has received funding from the European Research Council (ERC) under the European Union's Horizon 2020 research and innovation programme (grant agreement No 810367), project EMC2. AJM and JPP acknowledge funding from RSC (\verb!IEC\R2\181027!) and CNRS under the joint research project (PRC) grant. Computations have been performed at GENCI on the Occigen machine (CINES, Montpellier, France) on grant no A0070707671. 
We acknowledge particular thanks to one of our referees for suggesting changes that have considerably increased the scientific content of this paper.

\bibliography{BIBLIOGRAPHY}



\end{document}

%% file: macros.tex
\usepackage{xspace}

\newcommand{\etal}{\emph{et al.}\xspace}
\newcommand{\abinitio}{{\em ab initio}\xspace}

\newcommand{\secrff}[1]{{\ \S~\ref{#1}}}
\newcommand{\figrff}[1]{Figure~\ref{#1}}
\newcommand{\tabrff}[1]{Table~\ref{#1}}
\newcommand{\eqrff}[1]{eq.~\eqref{#1}}

\newcommand{\subfigrff}[2]{Fig.~\ref{#1}{#2}}

\newcommand{\ELST}[0]{\ensuremath{E_{\rm elst}}\xspace}
\newcommand{\EXCH}[0]{\ensuremath{E_{\rm exch}}\xspace}

\newcommand{\EpolSAPT}[1]{\ensuremath{E^{(#1)}_{\mathrm{pol}}}\xspace}
\newcommand{\EexchSAPT}[1]{\ensuremath{E^{(#1)}_{\mathrm{exch}}}\xspace}

\newcommand{\Eelst}[0]{\ensuremath{E^{(1)}_{\rm elst}}\xspace}

\newcommand{\Eexch}[0]{\ensuremath{E^{(1)}_{\rm exch}}\xspace}

\newcommand{\EINDreg}[1]{\ensuremath{E^{(#1)}_{\rm IND}({\rm Reg})}\xspace}
\newcommand{\EIND}[1]{\ensuremath{E^{(#1)}_{\rm IND}}\xspace}
\newcommand{\Eindpol}[1]{\ensuremath{E^{(#1)}_{\rm ind}}\xspace}

\newcommand{\Eexind}[1]{\ensuremath{E^{(#1)}_{\rm ind,exch}}\xspace}

\newcommand{\EDISP}[1]{\ensuremath{E^{(#1)}_{\rm DISP}}\xspace}
\newcommand{\Edisppol}[1]{\ensuremath{E^{(#1)}_{\rm disp}}\xspace}
\newcommand{\Eexdisp}[1]{\ensuremath{E^{(#1)}_{\rm disp,exch}}\xspace}

\newcommand{\EINT}[0]{\ensuremath{E_{\rm int}}\xspace}
\newcommand{\Ecomp}[1]{\ensuremath{E_{\mathrm{#1}}}\xspace}

\newcommand{\Eint}[1]{\ensuremath{E_{\rm int}^{[#1]}}\xspace}

\newcommand{\sEDISP}[1]{\ensuremath{E_{\rm DISP}^{[#1]}}\xspace}

\newcommand{\Ssq}{\ensuremath{S^2}\xspace}

\newcommand{\noSsq}{\ensuremath{\text{no}S^2}\xspace}

\newcommand{\DHF}[0]{\ensuremath{\delta^{\rm HF}_{\rm int}}\xspace}
\newcommand{\dHF}[1]{\ensuremath{\delta^{\rm HF}_{\rm int}[#1]}\xspace}
\newcommand{\DELTAHF}[1]{\ensuremath{\dHF{#1}}\xspace}

\newcommand{\Epol}[1]{\ensuremath{E_{\rm POL}^{(#1)}}\xspace}
\newcommand{\EPOL}[0]{\ensuremath{E_{\rm POL}}\xspace}

\newcommand{\EpolCL}[1]{\ensuremath{E^{(#1)}_{\mathrm{pol,cl}}}\xspace}
\newcommand{\EPOLCL}[0]{\ensuremath{E_{\mathrm{pol,cl}}}\xspace}

\newcommand{\Ecd}[1]{\ensuremath{E^{(#1)}_{\rm CD}}\xspace}
\newcommand{\EcdSM}[1]{\ensuremath{E^{(#1)}_{\rm CD}({\rm SM09})}\xspace}
\newcommand{\EcdREG}[1]{\ensuremath{E^{(#1)}_{\rm CD}(\rm Reg)}\xspace}
\newcommand{\ECD}[0]{\ensuremath{E_{\rm CD}}\xspace}
\newcommand{\ecd}[0]{\ensuremath{E_{\rm cd}}\xspace}
\newcommand{\EintHF}[0]{\ensuremath{E_{\rm int}^{\rm HF}}\xspace}

\newcommand{\EintSAPT}[1]{\ensuremath{ E_{\mathrm{int}}^{\mathrm{SAPT{#1}}} }\xspace}

\newcommand{\EintSAPTDFT}[0]{\ensuremath{ E_{\mathrm{int}}^{\mathrm{SAPT(DFT)}} }\xspace}
\newcommand{\EELEC}[2]{\ensuremath{E_{\mathrm{elst}}^{(#1#2)}}\xspace}
\newcommand{\EEXCH}[2]{\ensuremath{E_{\mathrm{exch}}^{(#1#2)}}\xspace}
\newcommand{\EINDD}[2]{\ensuremath{E_{\mathrm{ind,r}}^{(#1#2)}}\xspace}
\newcommand{\EEXCHIND}[2]{\ensuremath{E_{\mathrm{ind,exch,r}}^{(#1#2)}}\xspace}

\newcommand{\EDISPP}[2]{\ensuremath{E_{\mathrm{disp}}^{(#1#2)}}\xspace}
\newcommand{\EEXCHDISP}[2]{\ensuremath{E_{\mathrm{disp,exch}}^{(#1#2)}}\xspace}
\newcommand{\EINDDISP}[2]{\ensuremath{E_{\mathrm{ind-disp}}^{(30)}}\xspace}
\newcommand{\EEXCHINDDISP}[2]{\ensuremath{E_{\mathrm{ind-disp,exch}}^{(30)}}\xspace}

\newcommand{\pbez}{{PBE0}\xspace}
\newcommand{\wbxd}{{$\omega$B97X-D}\xspace}
\newcommand{\blyp}{{B3LYP}\xspace}

\newcommand{\almo}{{ALMO}\xspace}
\newcommand{\almofunc}[1]{{{#1}||ALMO}\xspace}
\newcommand{\rvshf}{{HF||RVS}\xspace}
\newcommand{\csovblyp}{{B3LYP||CSOV}\xspace}

\newcommand{\sapt}{{SAPT}\xspace}
\newcommand{\SAPT}[1]{{SAPT#1}\xspace}
\newcommand{\saptdft}{{SAPT(DFT)}\xspace}

\newcommand{\regsaptdft}{{Reg-SAPT(DFT)}\xspace}

\newcommand{\saptdftpbezssq}{{SAPT(DFT)[PBE0/AC/\ensuremath{{S^2}}]}\xspace}
\newcommand{\saptdftssq}{{SAPT(DFT)/\ensuremath{{S^2}}}\xspace}
\newcommand{\saptdftnossq}{{SAPT(DFT)/no\ensuremath{{S^2}}}\xspace}
\newcommand{\ccsdt}{{CCSD(T)}\xspace}

\newcommand{\dft}{{DFT}\xspace}

\newcommand{\rvs}{{RVS}\xspace}
\newcommand{\csov}{{CSOV}\xspace}

\newcommand{\amoeba}{ {AMOEBA}\xspace}

\newcommand{\polFF}{{polFF}\xspace}

\newcommand{\Q}[2]{\ensuremath{Q^{#1}_{#2}}\xspace}  

\newcommand{\T}[2]{\ensuremath{T^{#1}_{#2}}\xspace}
\newcommand{\A}[2]{\ensuremath{\alpha^{#1}_{#2}}\xspace}

\newcommand{\BETApol}[1]{\ensuremath{\beta_{\rm pol}^{#1}}\xspace}

\newcommand{\CamCASP}{{\sc CamCASP}\xspace}
\newcommand{\PsiF}{{\sc Psi4}\xspace}

\newcommand{\Gamess}{{\sc GAMESS(US)}\xspace}
\newcommand{\Gaussian}{{\sc Gaussian09} \xspace}
\newcommand{\Molpro}{{\sc Molpro}\xspace}

\newcommand{\Hondo}{{\sc Hondo} \xspace}
\newcommand{\Qchem}{{\sc Qchem} \xspace}

\newcommand{\water}{H\textsubscript{2}O\xspace}
\newcommand{\Zn}{Zn\textsuperscript{$2+$}\xspace}

\newcommand{\F}{F\textsuperscript{$-$}\xspace}
\newcommand{\Cl}{Cl\textsuperscript{$-$}\xspace}
\newcommand{\OH}{OH\textsuperscript{$-$}\xspace}
\newcommand{\waterdimer}{(\water)$_2$\xspace}

\newcommand{\Znwater}{\Zn\!\!\dots\water}

\newcommand{\Fwater}{\F\!\!\dots\water}
\newcommand{\Clwater}{\Cl\!\!\dots\water\xspace}
\newcommand{\OHwater}{\OH\!\!\dots\water\xspace}

\newcommand{\kcalmol}{kcal mol$^{-1}$\xspace}

\newcommand{\SI}{{SI}\xspace}

%% file: Structures-Representation.tex
\begin{figure}[H]
\captionsetup{justification=raggedright,singlelinecheck=false}
\subfloat[]{\includegraphics[scale=0.25]{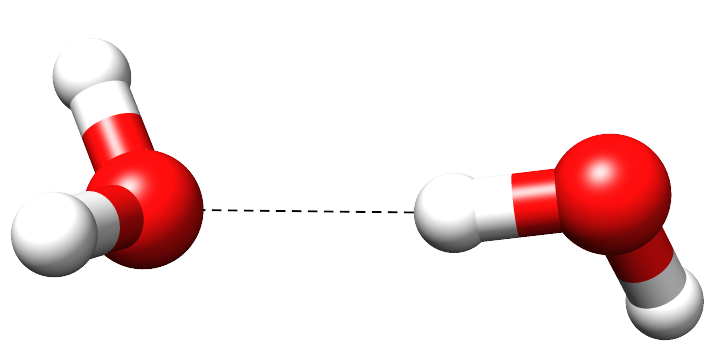}\label{figwater}}
\hspace*{.3cm}
\subfloat[]{\includegraphics[scale=0.18]{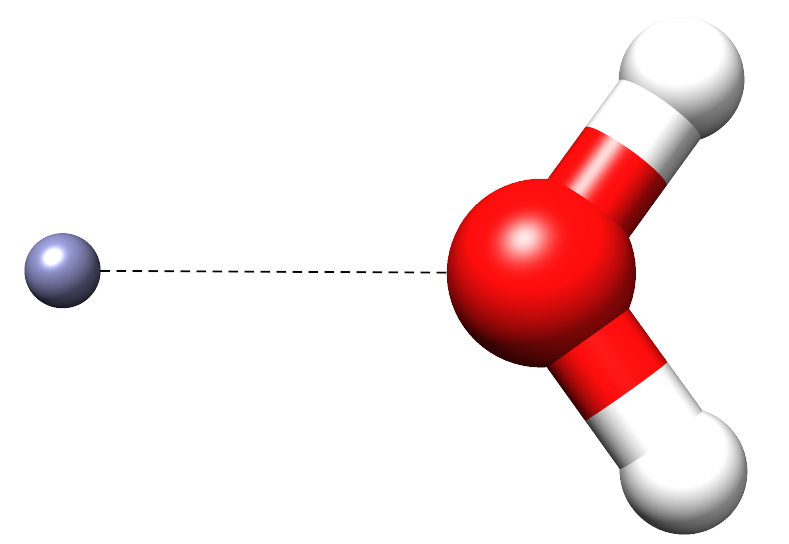}\label{figzinc}} 
\hspace*{.3cm}
\subfloat[]{\includegraphics[scale=0.22]{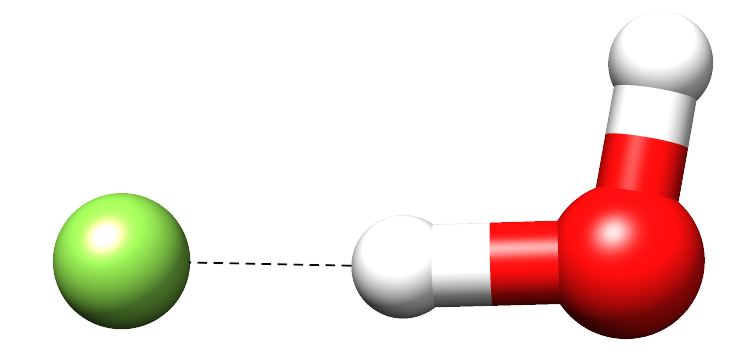}\label{figfluor}}



\caption{Graphical representation of \protect\subref{figwater} \water\dots\water, \protect\subref{figzinc} \Zn\dots\water and \protect\subref{figfluor} \F\dots\water complexes.
}
\label{fig:structures}
\end{figure}

%% file: table-saptorder.tex
\begin{landscape}
\captionsetup[table]{singlelinecheck=false} 
\begin{table}
\begin{flushleft}  
\begin{tabular}{lccc}
\arrayrulecolor{blue}
\textbf{Electrostatics} \\
\toprule
\SAPT{0} & \cellcolor{myblue}\EELEC{1}{0} & & \\
\SAPT{2} & " & \cellcolor{myblue}+\EELEC{1}{2} & \\
\SAPT{2+} & " & " & \\
\SAPT{2+(3)} & " & " & \cellcolor{myblue}+\EELEC{1}{3} \\
\SAPT{2+3} & " & " & " \\
\bottomrule
\end{tabular} 
\hspace{0.5cm}
\begin{tabular}{rrr}
\arrayrulecolor{red}
\multicolumn{3}{l}{\textbf{Exchange-Repulsion}} \\
\toprule
\cellcolor{myred}+\EEXCH{1}{0} & & \\
" & \cellcolor{myred}+\EEXCH{1}{1} & \cellcolor{myred}+\EEXCH{1}{2} \\
" & " & " \\
" & " & " \\
" & " & " \\
\bottomrule
\end{tabular} 
\newline
\vspace{0.5cm}
\newline
\begin{tabular}{lccccccc}
\arrayrulecolor{green}
\textbf{Induction} \\
\toprule
\SAPT{0} & \cellcolor{mygreen}+\dHF{2} & \cellcolor{mygreen}+\EINDD{2}{0} & \cellcolor{mygreen}+\EEXCHIND{2}{0} &  & & & \\
\SAPT{2} & " & " & " & +$^{t}$\cellcolor{mygreen}\EINDD{2}{2} & +$^{t}$\cellcolor{mygreen}\EEXCHIND{2}{2} & &\\
\SAPT{2+} & " & " & " & " & " & &  \\
\SAPT{2+(3)} & " & " & " & " & " & &  \\
\SAPT{2+3} & \cellcolor{mygreen}$+\dHF{3}-\dHF{2}$ & " & " & " & " &  \cellcolor{mygreen}+\EINDD{3}{0} & \cellcolor{mygreen}+\EEXCHIND{3}{0} \\
\bottomrule
\end{tabular}
\newline
\vspace{0.5cm}
\newline
\begin{tabular}{lcccccccc}
\arrayrulecolor{violet}
\textbf{Dispersion} \\
\toprule
\SAPT{0} & \cellcolor{mypurple}+\EDISPP{2}{0} & \cellcolor{mypurple}+\EEXCHDISP{2}{0} &  &  & & & & \\
\SAPT{2} & " & " &  &  &  & & & \\
\SAPT{2+} & " & " & \cellcolor{mypurple}+\EDISPP{2}{1} & \cellcolor{mypurple}\EDISPP{2}{2}  &  & & & \\
\SAPT{2+(3)} & " & " & " & " &  \cellcolor{mypurple}+\EDISPP{3}{0} & & &  \\
\SAPT{2+3} & " & " & " & " & " &  \cellcolor{mypurple}+\EEXCHDISP{3}{0} & \cellcolor{mypurple}+\EINDDISP{3}{0} & \cellcolor{mypurple}+\EEXCHINDDISP{3}{0} \\
\bottomrule
\end{tabular}
\end{flushleft}
\caption{Summary of commonly used SAPT methods.
}
\label{tabsapt}
\end{table}
\end{landscape}

%% file: Eint_water_zn.tex
\captionsetup{justification=justified,singlelinecheck=false}

\begin{figure}

\begin{center}

   \includegraphics[viewport=60 90 1000 1190, clip, scale=0.48]{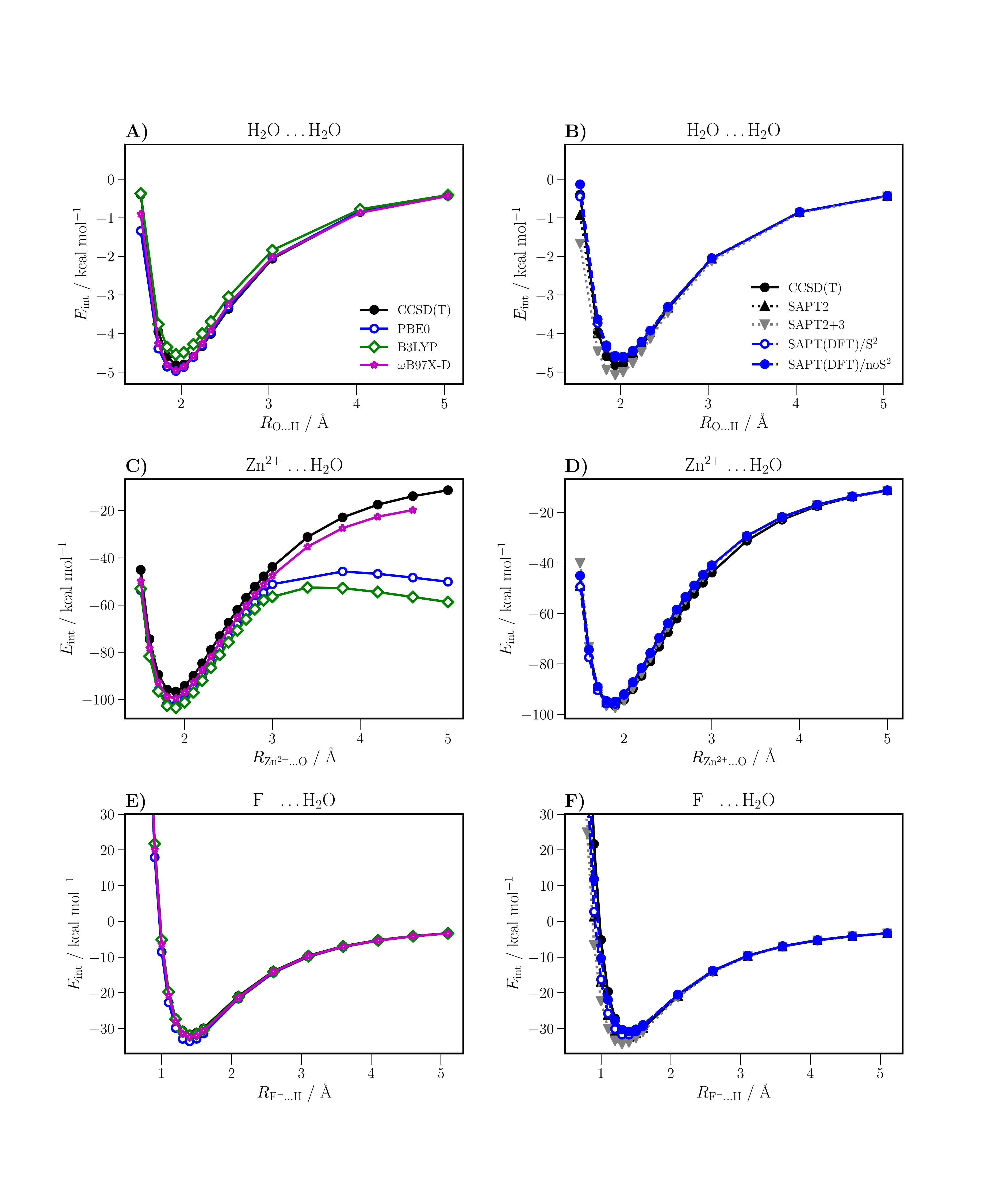}


\caption{ 
Comparison of the intermolecular interaction energy between DFT-based supermolecular EDAs, \sapt models, \saptdft and \ccsdt for the water dimer, \Zn\dots\water and \F\dots\water complexes. The asymptotically corrected PBE0 functional is used for \saptdft. The \Ssq approximation is used for the second and third-order exchange energies in the \sapt models and \saptdftssq. For \saptdftnossq this approximation is present only in the \Eexdisp{2} energy.
}

\label{fig:Zn-H2O-Eint}
\end{center}
\end{figure}

%% file: SAPT_SAPTdft_and_dHF_S2_and_noS2.tex
\begin{figure}[H]

\begin{center}

  \includegraphics[viewport=60 90 1000 1190,clip,scale=0.50]{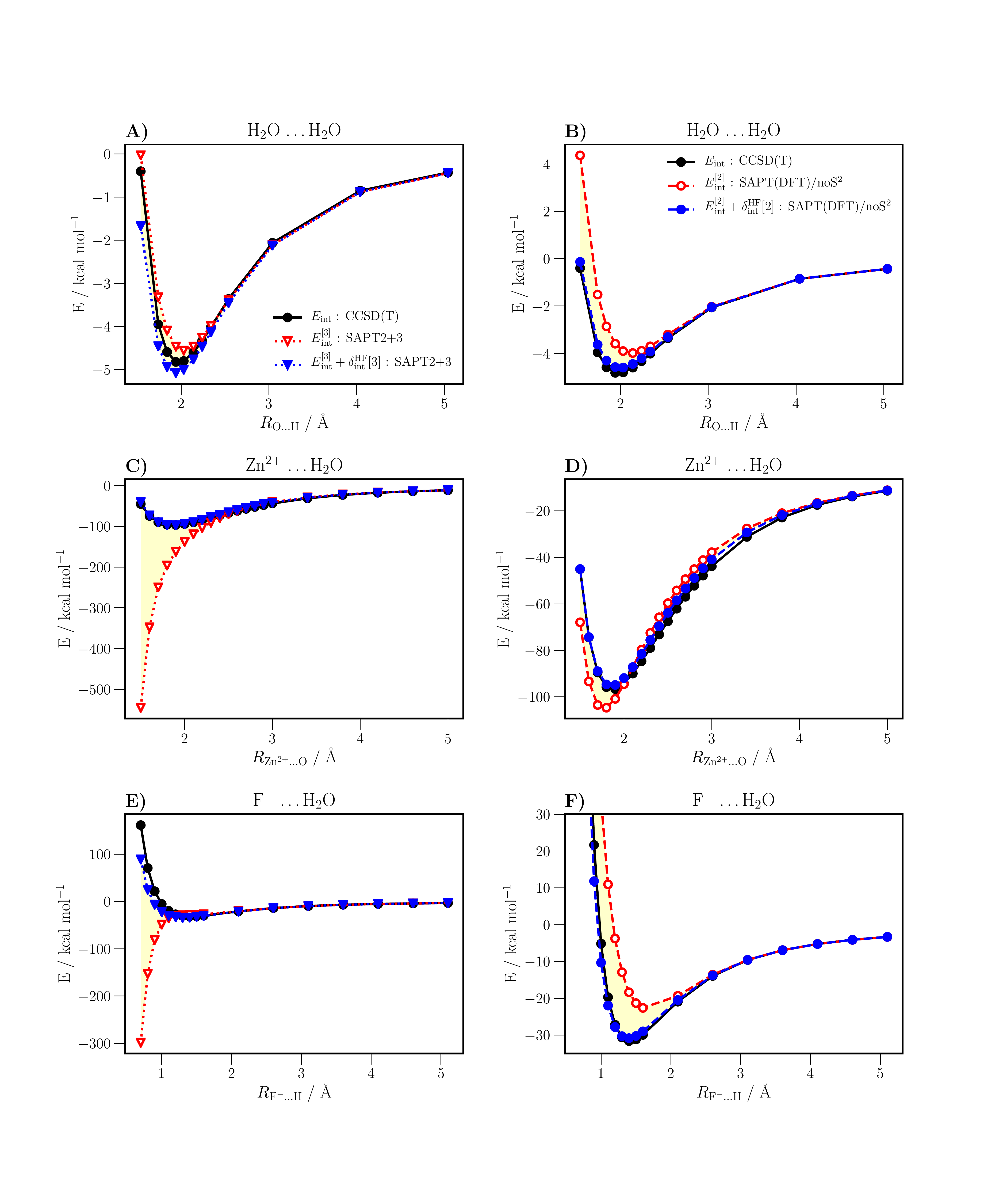}
\end{center}
\caption{ 
  Contribution of the \DHF energy to the interaction energies from \SAPT{2+3} and \saptdft. The asymptotically corrected PBE0 functional is used for \saptdft. The \Ssq approximation is used for the second and third-order exchange energies in \SAPT{2+3}. For \saptdft this approximation is present only in the \Eexdisp{2} energy. For \saptdft the difference between the \Eint{2} and $\Eint{2}+\dHF{2}$ curves is the \dHF{2} contribution, and for \SAPT{2+3} the difference between the \Eint{3} and  $\Eint{3}+\dHF{3}$ curves is the \dHF{3} contribution. These regions are shaded yellow.
}
\label{fig:SAPT_SAPTDFT_and_dHF_S2_and_noS2}

\end{figure}

%% file: Comparison_of_SAPT_SAPTDFT_and_EDAs.tex

\begin{figure}

  
  \includegraphics[viewport=60 90 1000 1190, clip, scale=0.50]{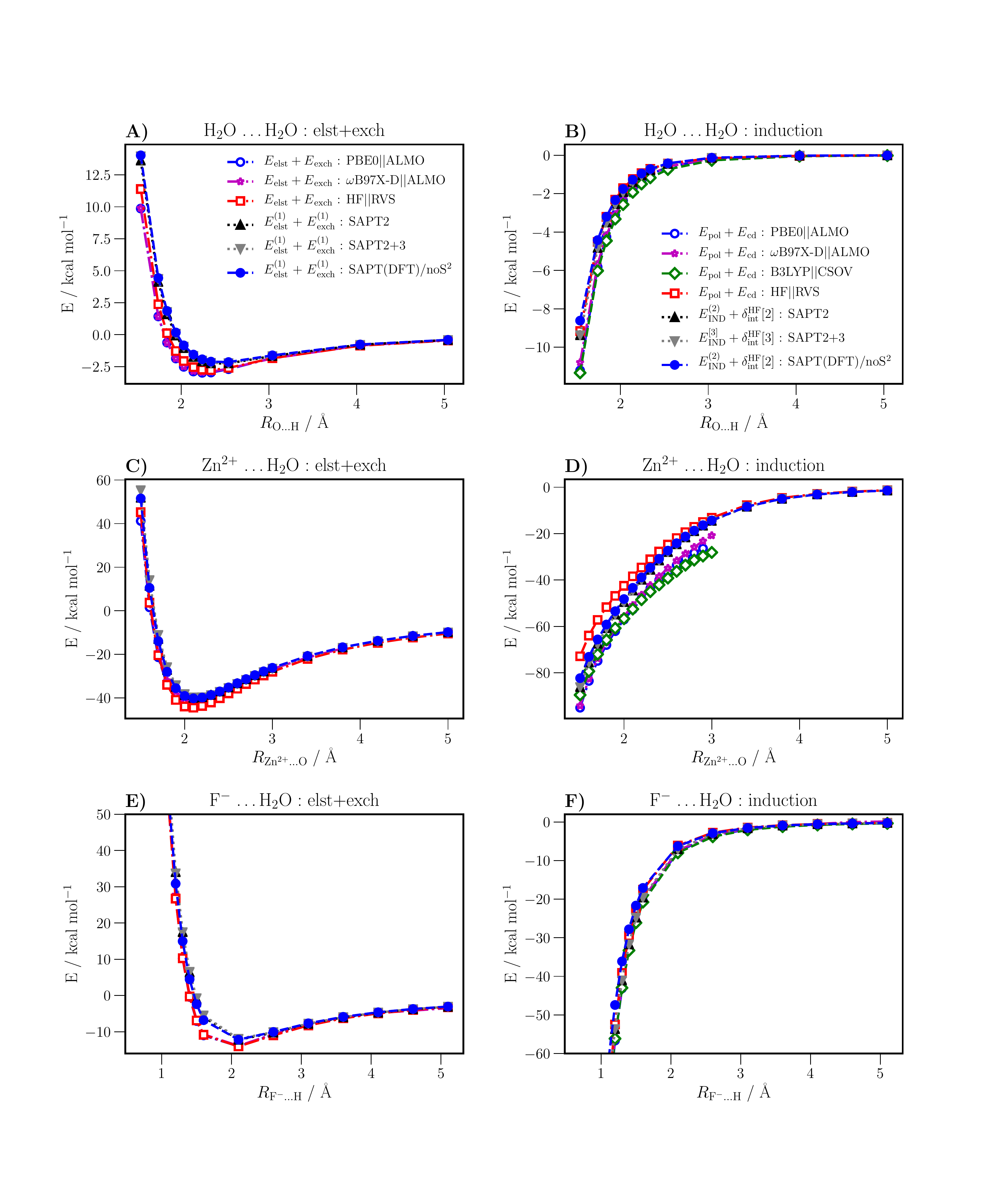}
  


\caption{ 
  Comparison of the main contributions from \sapt models and \saptdft with those from supermolecular EDAs for the water dimer, \Zn\dots\water and \F\dots\water complexes. The asymptotically corrected PBE0 functional is used for \saptdft. In the \sapt models, the \Ssq approximation is used for the second and third-order exchange energies. For \saptdft this approximation is present only in the \Eexdisp{2} energy.
}
\label{fig:Comparison_of_SAPT_SAPTDFT_and_EDAs}

\end{figure}

%% file: CD_POL_from_SAPTDFT_and_EDAs.tex
\begin{figure}[H]

\includegraphics[viewport=60 90 1000 1190, clip, scale=0.48]{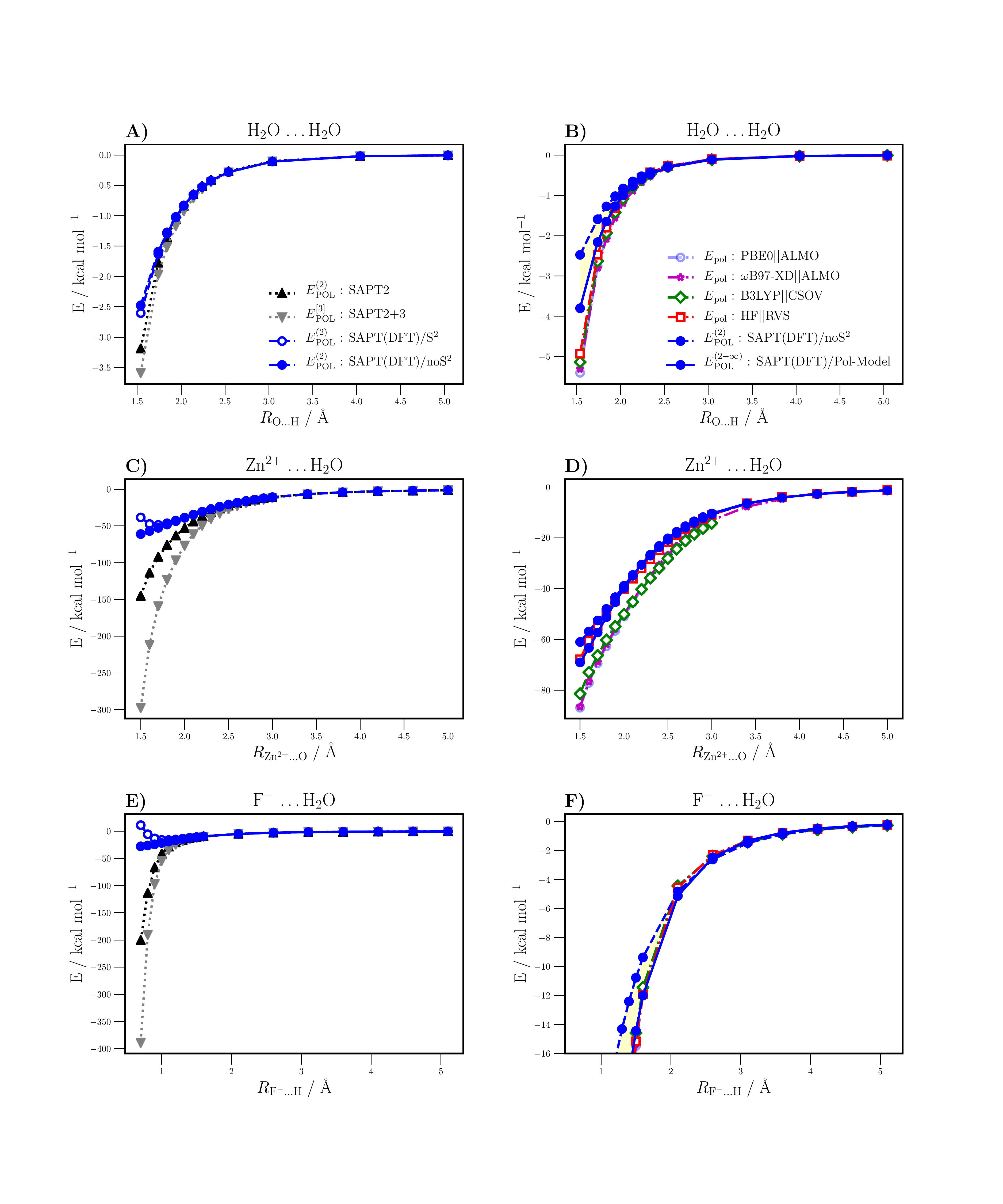}

\caption{ 
  Comparison of \EPOL from SAPT and the supermolecular EDAs for the water dimer, \Zn\dots\water and \F\dots\water complexes. 
  The \Ssq approximation is used for the second and third-order exchange energies in the \sapt models and \saptdftpbezssq. For \saptdftnossq this approximation is removed from the energies shown here. And \Epol{2-\infty}: \saptdft/Pol-Model is obtained with the polarization model developed in this work.
}
\label{fig:POL_from_SAPTDFT_and_EDAs}

\end{figure}

\begin{figure}

\includegraphics[viewport=60 90 1000 1190, clip, scale=0.48]{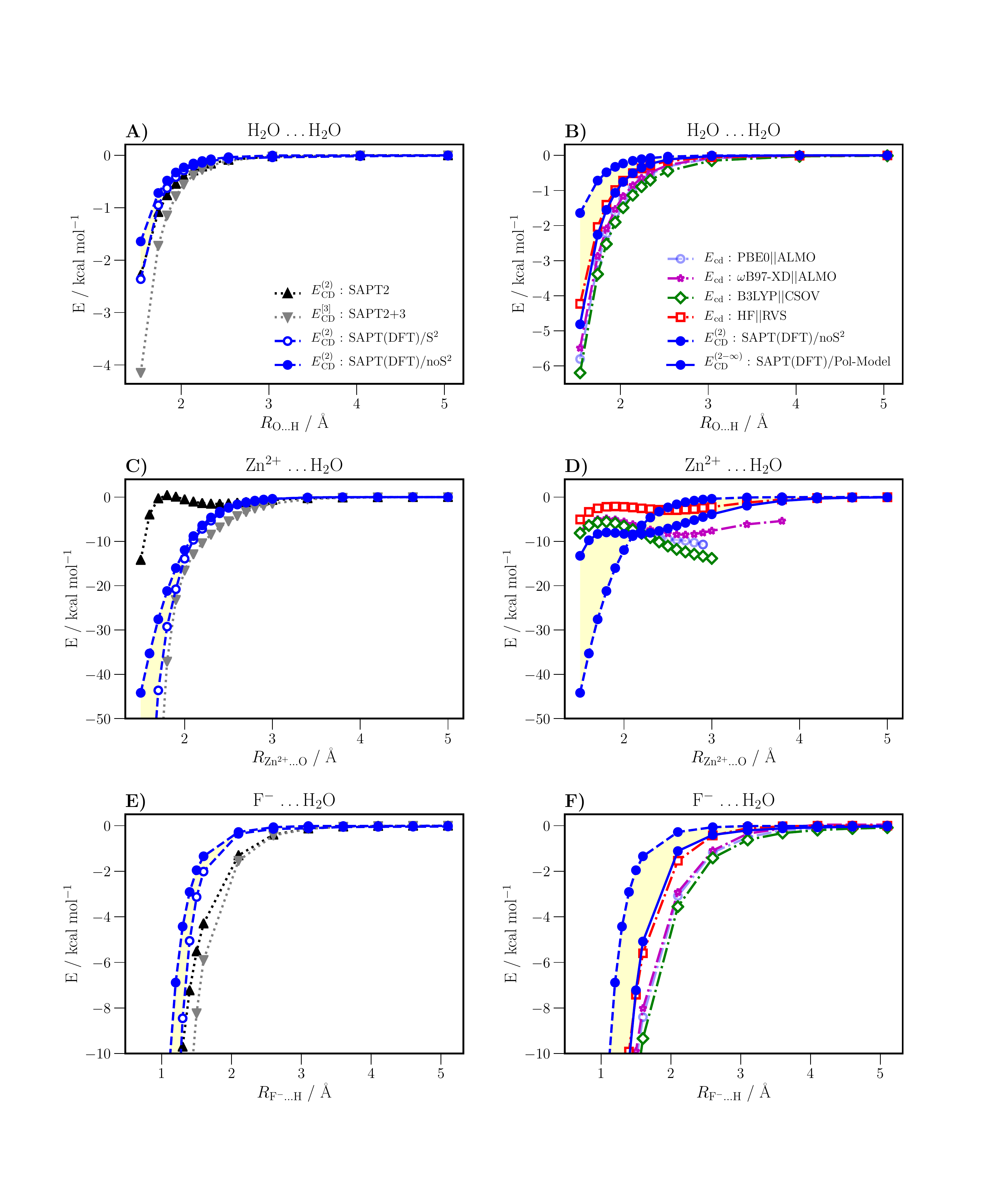}

\caption{ 
  Comparison of \ECD from SAPT and the supermolecular EDAs for the water dimer, \Zn\dots\water and \F\dots\water complexes.
  The \Ssq approximation is used for the second and third-order exchange energies in the \sapt models and \saptdftssq. For \saptdftnossq this approximation is removed from the energies shown here.  \Ecd{2-\infty}: \saptdft/Pol-Model is obtained with the polarization model developed in this work. For \Fwater calculations with \almofunc{\wbxd} there were numerical issues at the two longest separations and so these have been removed from panel (F) in the figure.
}
\label{fig:CD_from_SAPTDFT_and_EDAs}

\end{figure}

%% file: table-notation.tex
\begin{table}[H]
\begin{tabular}{cl}
\toprule
\midrule
Notation & Definition \\
\midrule
 & X =  \\
 & \tabitem elst: electrostatic   \\
 & \tabitem exch: exchange-repulsion   \\
$E_{\rm X}$ & \tabitem pol/POL: polarization (from EDAs/SAPTs)  \\
 & \tabitem cd/CD: charge-delocalization (from EDAs/SAPTs)   \\
 & \tabitem IND: sum of induction and exchange-induction   \\
 & \tabitem DISP: sum of dispersion and exchange-dispersion   \\
 & \tabitem inter: total intermolecular interaction  \\
 
\midrule

$E_{\rm X}^{\rm (n)}$ & contribution X of order (n) = 1, 2 or 3  \\
\\
$E_{\rm X}^{\rm [n]}$ & sum of contribution X of order up to [n] = 2 or 3  \\
\\
\dHF{n} & delta Hartree-Fock term of order n = 2 or 3 \\

\midrule

& \tabitem F = \blyp, \pbez, \wbxd \\
$E_{\rm X}$: F$||$M  & \tabitem M = \ccsdt, \almo, \csov, \rvs  \\ 
\\
$E_{\rm X}$: N &  N = \SAPT{2}, \SAPT{2+3}, \saptdft \\
\midrule
\saptdftssq & \Eexind{2} computed with \Ssq \\
\saptdftnossq & \Eexind{2} computed without \Ssq \\
& \Ssq = Single-Exchange Approximation \\

\midrule
& Y = POL, CD  \\
$E^{(2/3)}_{\rm Y}$: \SAPT{2}/\SAPT{2+3} & from the SM09 CD Definition \\
$E^{(2)}_{\rm Y}$: \saptdft & from the Reg-CD Definition \\
$E^{(2-\infty)}_{\rm Y}$: \saptdft/Pol-Model & from the polarization model developed in this work \\

\midrule
\bottomrule
\end{tabular}
\caption*{List of symbol used in this work.}
\label{tabnot}
\end{table}